\documentclass[12pt]{article}
\usepackage{a4wide,epsfig,psfrag,amsmath,amssymb,cite,scalefnt}
\usepackage{color}
\usepackage{amsmath,comment,braket}
\usepackage{placeins}
\usepackage{dsfont}
\usepackage{multirow}
\usepackage{rotating} 

\graphicspath{ {figs/} }

\newcommand{\be}{\begin{equation}}
\newcommand{\ee}{\end{equation}}
\newcommand{\bea}{\begin{eqnarray}}
\newcommand{\eea}{\end{eqnarray}}

\allowdisplaybreaks

\newcommand{\gsim}{\;\rlap{\lower 3.5 pt \hbox{$\mathchar \sim$}} \raise 1pt
 \hbox {$>$}\;}
\newcommand{\lsim}{\;\rlap{\lower 3.5 pt \hbox{$\mathchar \sim$}} \raise 1pt
 \hbox {$<$}\;}

\begin{document}

\title{\vskip-3cm{\baselineskip14pt
    \begin{flushleft}
      \normalsize P3H-19-023\\
      \normalsize TTP19-024
  \end{flushleft}}
  \vskip1.5cm
  Matching coefficients in NRQCD to two-loop accuracy
}

\author{
  Marvin Gerlach$^{a}$,
  Go Mishima$^{a,b}$,
  Matthias Steinhauser$^{a}$
  \\[1mm]
  {\small\it $^a$Institut f{\"u}r Theoretische Teilchenphysik}\\
  {\small\it Karlsruhe Institute of Technology (KIT)}\\
  {\small\it Wolfgang-Gaede Stra\ss{}e 1, 76128 Karlsruhe, Germany}
  \\[1mm]
  {\small\it $^b$Institut f{\"u}r Kernphysik}\\
  {\small\it Karlsruhe Institute of Technology (KIT)}
  \\
  {\small\it Hermann-von-Helmholtz-Platz 1, 76344 Eggenstein-Leopoldshafen, Germany}
}
  
\date{}

\maketitle

\thispagestyle{empty}

\begin{abstract}

  We consider the Lagrange density of non-relativistic Quantum Chromodynamics
  expanded up to order $1/m^2$, where $m$ is the heavy quark mass, and compute
  several matching coefficients up to two-loop order. Our results are building
  blocks for next-to-next-to-next-to-leading logarithmic and 
  next-to-next-to-next-to-next-to-leading order corrections to the
  threshold production of top quark pairs and the decay of heavy quarkonia.
  We describe the techniques used for the calculation and provide
  analytic results for a general covariant gauge.
  
%

\end{abstract}

\thispagestyle{empty}

\sloppy


\newpage


\section{\label{sec::intro}Introduction}

Non-Relativistic Quantum Chromodynamics (NRQCD)~\cite{Bodwin:1994jh}
has proven to provide accurate predictions for systems of two heavy
quarks, which move with a small relative velocity.  Among them are
decay rates and binding energies of quarkonia and the threshold
production of top quark pairs in electron positron annihilation.  For
comprehensive compilations of results we refer to the review
articles~\cite{Brambilla:2004jw,Pineda:2011dg,Beneke:2013jia} and
restrict ourselves here to recent next-to-next-to-next-to-leading
order (N$^3$LO) results.  These include predictions for top quark pair
production~\cite{Beneke:2015kwa},\footnote{ In Ref.~\cite{Hoang:2013uda}
next-to-next-to-leading logarithmic (NNLL) corrections have been
obtained, see also~\cite{Pineda:2011aw}.} the decay of the $\Upsilon(1S)$ meson~\cite{Beneke:2014qea},
and energy levels of heavy quarkonia ground 
and excited states~\cite{Penin:2002zv,Kiyo:2013aea,Peset:2018ria}
together with  phenomenological applications~\cite{Kiyo:2015ufa,Mateu:2017hlz}.

Despite the high accuracy reached for a number of observables, it is
desirable to extend the precision of the predictions.  For example, the
perturbative uncertainty of the N$^3$LO top quark threshold prediction of about
3\% will constitute the main uncertainty in the top quark mass value 
extracted from the comparison with future cross section measurements (see,
e.g., Ref.~\cite{Simon:2016pwp}).  Furthermore, the dominant source of
uncertainty in the determination of the charm and bottom quark masses
from bound state energies
originates from the renormalization scale dependence, due to unknown higher
order corrections~\cite{Kiyo:2015ufa,Peset:2018ria}.  Currently a complete
N$^4$LO calculation is out of reach, note, however, that the completion of
the ingredients necessary for the N$^3$LO predictions took more than ten years
and the combined effort of several groups (see, e.g.,
Ref.~\cite{Beneke:2013jia}).  It is thus reasonable to proceed in a similar way
at N$^4$LO and gradually provide the individual building blocks required. In this work
we compute two-loop matching coefficients which are building blocks of the
NRQCD Lagrange density at N$^4$LO.

A further and more short-term motivation of our work is the
construction of logarithmically enhanced contributions which
complement the N$^3$LO predictions.  The potential NRQCD (pNRQCD) Lagrange density
relevant for $S$-wave states with next-to-next-to-next-to-leading
logarithmic (N$^3$LL) accuracy has been constructed in
Ref.~\cite{Anzai:2018eua} up to a few missing contributions to the so-called
soft running. Among them are the coefficients $d_{ss}$ and $d_{vs}$
(see the next section for a precise definition)
which are computed in this work.
Note that for $P$-wave states the N$^3$LL pNRQCD Lagrange density is complete and
can be found in Ref.~\cite{Peset:2018jkf}.

The main purpose of this paper is the computation of the matching coefficients
between QCD and NRQCD to two-loop order. We concentrate on the four-fermion
operators but also compute the matching coefficients for gluon-quark
interactions ($c_D$, $c_F$ and $c_S$) which are needed to obtain gauge
invariant results.  The corresponding one-loop results have been obtained in
Refs.~\cite{Pineda:1998kj} and~\cite{Manohar:1997qy}, respectively (see also
Refs.~\cite{Beneke:2013jia}).  The gauge dependence has its origin in the
non-minimality of the operators entering the NRQCD Lagrange density. If fact, some
of the effective operators can be absorbed into other operators by using the
equation of motion or field redefinitions.  The relevant equation of motion  in our
calculation is that which relates some of the four-fermion
operators and the gluon-quark interaction~\cite{Bauer:1997gs} and thus only a
particular combination is gauge invariant (see, e.g.,
Ref.~\cite{Pineda:2001ra}).  In this paper, we perform our calculations in the
general covariant gauge and present results for an arbitrary gauge parameter
$\xi$. We check the cancellation of $\xi$ in the proper combination of
the matching coefficients entering physical quantities.
The computation of $d_{xy}$ requires a precise definition of the Pauli
matrices in $d=4-2\epsilon$ dimensions, which we discuss in detail.

\begin{figure}[t]
  \centering
      \includegraphics[width=\textwidth]{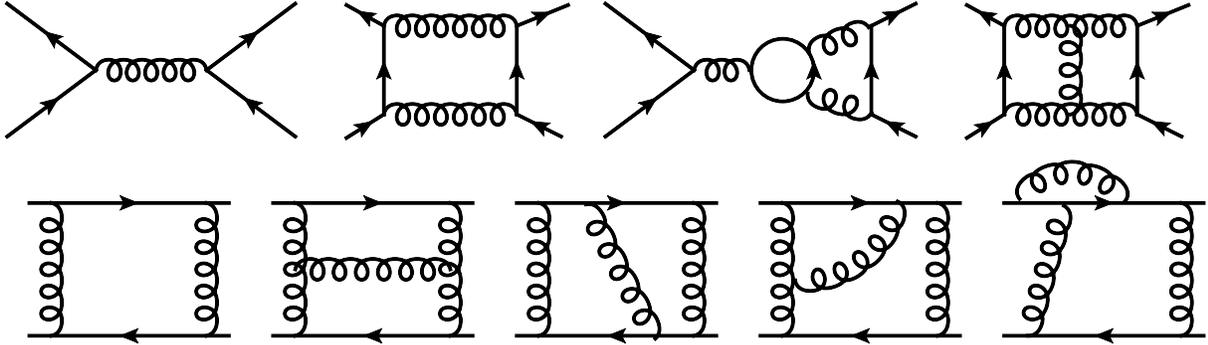}
  \caption{\label{fig::diags} Sample Feynman diagrams contributing to
    $d_{xy}$.}
\end{figure}

The calculation of the matching coefficients for four-fermion operators is
naturally divided into two parts, which we call the annihilation and the
scattering channel. The tree-level contribution of the former
originates from the diagrams where a quark-anti-quark pair annihilates into a
(virtual) gluon which subsequently ``decays'' into a quark-anti-quark pair
(cf. Fig.~\ref{fig::diags}).  The corresponding one- and two-loop sample diagrams
are shown in Figs.~\ref{fig::diags} and~\ref{fig::vector-current}.  In the
case of the scattering channel one considers the scattering of a quark and an
anti-quark, which may have different flavours and thus also different masses.

The remainder of the paper is organized as follows: In the next section we
provide the relevant parts of the NRQCD Lagrange density and define the matching
coefficients which we want to compute. In Section~\ref{sec::dxy}
we concentrate on the four-fermion matching coefficients and provide details
of our two-loop calculation. Section~\ref{sec::cD}
is devoted to the computation of the gluon fermion form factor and the
extraction of the corresponding matching coefficients.
The main results of the paper are presented in Section~\ref{sec::res}
where we provide analytic expressions for the four-fermion matching coefficients.
In the appendix we provide additional material such as the matching coefficients
needed for the redefinition of the gluon operators. Furthermore, analytic
results for all two-loop master integrals are given in Appendix~\ref{app::master}.

\section{\label{sec::lag}${\cal L}_{\rm NRQCD}$}

The NRQCD Lagrange density to order $1/m^2$ which 
we use for our calculations is given by (see, e.g., Refs.~\cite{Brambilla:2004jw,Beneke:2013jia})
\begin{align}
&\mathcal{L}_{\mathrm{NRQCD}}=\mathcal{L}_{g}+\mathcal{L}_{l}+\mathcal{L}_{\psi}+\mathcal{L}_{\chi}+\mathcal{L}_{\psi \chi}
\,,\\
&\mathcal{L}_{g}
=-\frac{1}{4} G^{\mu \nu a} G_{\mu \nu}^{a}
+\frac{1}{4} \frac{c_{1}^{g}}{m^{2}} g f_{a b c} G_{\mu \nu}^{a} G_{\alpha}^{\mu b} G^{\nu \alpha c}
\label{eq::lagGG}
\,,\\
&\mathcal{L}_{l}=\sum_{i=1}^{n_{l}} \overline{q}_{i} i /\!\!\!\! D q_{i}+\mathcal{O}\left(\frac{1}{m^{2}}\right)
\,,\\
& {\cal L}_{\psi}
 =
      \psi^\dagger\Bigg[
      iD_0 + \frac{c_k}{2m} \vec{D}^2 
      + g_s \frac{c_F}{2m} \vec{\sigma}\cdot \vec{B} 
      + g_s \frac{c_D}{8m^2} (\vec{D}\cdot \vec{E}-\vec{E}\cdot \vec{D}) 
      \nonumber\\
      &
      \qquad \qquad
      + i g_s \frac{c_S}{8m^2} \vec{\sigma} \cdot (\vec{D}\times \vec{E}-\vec{E}\times \vec{D}) 
      +\mathcal{O}\left(\frac{1}{m^3}\right)
      \Bigg]\psi
      \label{eq::LgQ}
  \,,\\
      &\mathcal{L}_{\chi} = - \mathcal{L}_\psi
      \quad \mathrm{with}
      \quad
      \psi \to \chi,
      iD^0\to -iD^0,
      E^i\to -E^i
      \,,
\end{align}
where $i \vec{D}=i\vec{\nabla} + g_s \vec{A}$, $E^j =G^{j0}$,
$B^j = -\varepsilon_{jkl} G^{kl}/2$, with $G^{ij}$ being the field strength
tensor, and $n_l$ is the number of light quarks.
In order to arrive at the canonical kinetic term of the gluon \eqref{eq::lagGG},
one has to apply the field redefinition and the rescaling~\cite{Pineda:2000sz}
(see also Appendix~\ref{app::d_i}).
 The main  purpose of this work is the computation of the matching
  coefficients of ${\cal L}_{\psi\chi}$ (see below). However, in order to
  construct a gauge invariant combination we also need $c_D$, which
  we discuss in Section~\ref{sec::cD}. Results for $c_F$
  and $c_S$ are presented in Appendix~\ref{app::cFcS}.

The interaction of four heavy quarks is given by 
\begin{eqnarray}
  {\cal L}_{\psi\chi}
  &=&
        \frac{d_{ss}}{m_1m_2} \psi_1^\dagger\psi_1\, \chi_2^\dagger\,\chi_2
      + \frac{d_{sv}}{m_1m_2} \psi_1^\dagger\vec{\sigma}\psi_1\, \chi_2^\dagger\vec{\sigma}\chi_2
      \nonumber\\&&\mbox{}
      + \frac{d_{vs}}{m_1m_2} \psi_1^\dagger T^a \psi_1\, \chi_2^\dagger T^a \chi_2
      + \frac{d_{vv}}{m_1m_2} \psi_1^\dagger T^a \vec{\sigma} \psi_1\, \chi_2^\dagger T^a \vec{\sigma} \chi_2
                    \,,
                    \label{eq::Ldxy}
\end{eqnarray}
where $\psi_1$ ($\psi_2$)
are Pauli spinors annihilating a heavy quark with mass $m_1$ ($m_2$), 
and $\chi_1$ ($\chi_2$) 
are Pauli spinors creating a heavy anti-quark with mass $m_1$ ($m_2$).
In this work we will identify the two masses and write
$m=m_1=m_2$. We furthermore use the notation for the subscripts which is
usually used in the literature: The first index in the matching coefficients
$d_{xy}$ refers to the colour (``s'' for singlet and ``v'' for octet) and the
second denotes the singlet (``s'') and triplet (``v'') quark-anti-quark
state.

The effective Lagrange density in Eq.~(\ref{eq::Ldxy}) can be rewritten with the
help of Fiertz transformations to arrive at
\begin{eqnarray}
 {\cal L}_{\psi\chi}
  &=&
        \frac{d^c_{ss}}{m_1m_2} \psi_1^\dagger\chi_2\, \chi_2^\dagger\,\psi_1
      + \frac{d^c_{sv}}{m_1m_2} \psi_1^\dagger\vec{\sigma}\chi_2\, \chi_2^\dagger\vec{\sigma}\psi_1
      \nonumber\\&&\mbox{}
      + \frac{d^c_{vs}}{m_1m_2} \psi_1^\dagger T^a \chi_2\, \chi_2^\dagger T^a \psi_1
      + \frac{d^c_{vv}}{m_1m_2} \psi_1^\dagger T^a \vec{\sigma} \chi_2\, \chi_2^\dagger T^a \vec{\sigma} \psi_1
                    \,,
                    \label{eq::Ldxy_2}
\end{eqnarray}
which is better suited for the annihilation part of the matching calculation
whereas we prefer version~(\ref{eq::Ldxy}) for the scattering part.
The relations between the coefficients in Eqs.~(\ref{eq::Ldxy})
and~(\ref{eq::Ldxy_2}) are given by~\cite{Pineda:1998kj}
\begin{eqnarray}
  d_{ss} &=& -\frac{d_{ss}^c}{2N_c} - \frac{3d_{sv}^c}{2N_c} 
             - \frac{N_c^2-1}{4N_c^2}d_{vs}^c
             - 3\frac{N_c^2-1}{4N_c^2}d_{vv}^c
             \,,\nonumber\\
  d_{sv} &=& -\frac{d_{ss}^c}{2N_c} + \frac{d_{sv}^c}{2N_c} 
             - \frac{N_c^2-1}{4N_c^2}d_{vs}^c
             + \frac{N_c^2-1}{4N_c^2}d_{vv}^c
             \,,\nonumber\\
  d_{vs} &=& -d_{ss}^c - 3d_{sv}^c + \frac{d_{vs}^c}{2N_c}
             + \frac{3d_{vv}^c}{2N_c}
             \,, \nonumber\\
  d_{vv} &=& -d_{ss}^c + d_{sv}^c  + \frac{d_{vs}^c}{2N_c}
             - \frac{d_{vv}^c}{2N_c}   
             \,,
\end{eqnarray}
where $N_c=3$ corresponds to QCD.
We compute the one- and two-loop four-quark amplitudes in Section~\ref{sec::dxy}
and provide results for $d_{xy}$ in Section~\ref{sec::res}.

Let us now describe the procedure which is used to obtain the
NRQCD matching coefficients.  We consider QCD with $n_h=1$ heavy quarks and
$n_l$ light quarks, and compute the four quark scattering amplitudes (see
Eqs.~\eqref{eq:m-scat} and~\eqref{eq:m-anni} below), the vertex corrections
(see Eq.~\eqref{eq::Gamma}), and the corrections to the  matching
  coefficients in the gluon sector (see Eq.~\eqref{eq::Lg}).  The
ultra-violet (UV) renormalization is done in the $(n_l+n_h)$-flavor theory.
The relation between the bare coupling constant $\alpha_s^0$ and
the $\overline{\mathrm{MS}}$ renormalized coupling constant
$\alpha_s(\mu)$ reads
\begin{align}
&\frac{\alpha_s^0}{\alpha_s(\mu)}
\left(\frac{\mu^2e^{\gamma_\mathrm{E}}}{4\pi}\right)^{-\epsilon}
=Z_{\alpha_s}
=1-\frac{\beta_0}{\epsilon}\frac{\alpha_s(\mu)}{\pi}
+\left(\frac{\beta_0^2}{\epsilon^2}-\frac{\beta_1}{2\epsilon}\right)
\left(\frac{\alpha_s(\mu)}{\pi}\right)^2
+\mathcal{O} (\alpha_s(\mu)^3)
\,,\\
&
\beta_{0}=\frac{11}{12} C_{A}-\frac{1}{3} (n_l+n_h) T_{F}, 
\quad \beta_{1}=\frac{17}{24} C_{A}^{2}-\left(\frac{5}{12} C_{A}+\frac{1}{4} C_{F}\right) (n_l+n_h) T_{F}
\,,
\end{align}
where $\mu$ is the renormalization scale,
and the colour factors for the $\mathrm{SU}(N_c)$ gauge group are given by
\begin{align}
T_{F}=\frac{1}{2}, \quad C_{F}=\frac{N_{c}^{2}-1}{2 N_{c}}, \quad C_{A}=N_{c}.
\end{align}
The heavy quark mass and wave function are renormalized on-shell.
The renormalization constants are well known in the literature
(see, e.g., Refs~\cite{Melnikov:2000zc,Marquard:2018rwx}).  We
recompute them here in order to retain the exact $\epsilon$-dependence.
Note that the wave function renormalization of the gluon 
is given by $1/\sqrt{Z_{\alpha_s}}$
because we use the background field method~\cite{Abbott:1980hw}.

We first compute
  $F_1^\prime(0)$, $F_2(0)$ (see Section~\ref{sec::cD}),
  and $d_1$, $d_2$ (see Appendix~\ref{app::d_i}).
After UV renormalization,
we convert the four-component Dirac spinors to 
the two-component Pauli spinors,
and 
the Dirac matrices $\gamma^\mu$ to
the Pauli matrices $\sigma^j$
assuming the non-relativistic limit.
We then canonicalize the gluon sector
(see Appendix~\ref{app::d_i}) and simultaneously
decouple the heavy quark in the gluon wave function.
 Finally,  we express 
$\alpha_s^{(n_l+n_h)}(\mu)=\alpha_s^{(n_l+1)}(\mu)$
in terms of $\alpha_s^{(n_l)}(\mu)$
by using the relation (for the bare version see Ref.~\cite{Grozin:2011nk})
\begin{align}
&\frac{\alpha_s^{(n_l+1)}(\mu)}{\alpha_s^{(n_l)}(\mu)}
=1-\frac{\alpha_s^{(n_l)}(\mu)}{\pi}
\frac{1-\epsilon I_0}{3\epsilon}T_F
+\left( \frac{\alpha_s^{(n_l)}(\mu)}{\pi} \right)^2 T_F
\left[ 
T_F
\frac{(1-\epsilon I_0)^2}{9\epsilon^2}
+C_A
\left(
-\frac{5}{24\epsilon}
\right.
\right.
\nonumber\\
& +
\left.
\frac{\epsilon  \left(4 \epsilon ^3+4 \epsilon ^2-11 \epsilon -10\right)I_0^2}
{8 (\epsilon -2) (2 \epsilon+1) (2 \epsilon +3)}
\right)
+
\left.
C_F\left(
\frac{-\epsilon  \left(4 \epsilon ^3-7 \epsilon -1\right)I_0^2}
{4 (\epsilon -2) (2 \epsilon -1) (2 \epsilon +1)}
-\frac{1}{8\epsilon}
\right)
\right]
+\mathcal{O}(\alpha_s^3)
\,,
  \label{eq:decoupling}
\end{align}
with $I_0=(\epsilon -1)I_1^a$, where $I_1^a$ is given in Eq.~\eqref{eq:MasterIntegrals}.
Equation~\eqref{eq:decoupling} is exact in $\epsilon$;
$\epsilon$-expanded versions can be found in Refs.~\cite{Chetyrkin:1997un,Grozin:2007fh}.
In order to keep the expressions in this paper simple
we provide the results in terms of $\alpha_s (m)$, which means that the
renormalization scale $\mu$ is set to $m$.  Using the renormalization group
equations it is possible to reexpress $\alpha_s(m)$ by $\alpha_s(\mu)$.
After expanding Eq.~(\ref{eq:decoupling}) in $\epsilon$ one obtains
$\log \mu^2/m^2$ terms which we abbreviate by
\begin{eqnarray}
  l_\mu &=&\log\frac{\mu^2}{m^2}
  \,.
\end{eqnarray}


\section{\label{sec::dxy}Four-fermion matching coefficients}

In this section we describe the calculation of the full-QCD amplitudes which
are needed for the matching coefficients $d_{xy}$ and $d_{xy}^c$ defined
in Eqs.~(\ref{eq::Ldxy}) and~(\ref{eq::Ldxy_2}).  They are obtained from
the four-quark amplitude
\begin{eqnarray}
  q_1(p) + \bar{q}_2(p) \to q_1(p) + \bar{q}_2(p)
  \label{eq::q1q2-q1q2}
\end{eqnarray}
with the special kinematics indicated in the arguments of the quark fields
$q_1$ and $q_2$.  Sample Feynman diagrams, which one has to consider at one-
and two-loop order, are shown in Fig.~\ref{fig::diags}.  In general one can
sub-divide them into ``annihilation'' (top row) and ``scattering''
contributions (bottom row).  Note that in the case that the two heavy quarks have
different flavours (and thus also different masses) only scattering diagrams contribute
whereas in the equal-mass case also the annihilation diagrams are needed. In
this paper we consider only the limit that both quarks have equal masses.
Nevertheless we discuss the two contributions separately.


\subsection{Matching}

Let us in the following briefly describe the individual steps which
are necessary to perform the matching between QCD and NRQCD. The
general idea is to consider the four-fermion amplitude in QCD in the
limit of a heavy quark mass and compare to the corresponding
expression in NRQCD, which provides results for $d_{xy}$ and $d_{xy}^c$.

We start with the QCD amplitudes which for the scattering and
annihilation channel have the form
\begin{align}
\mathcal{M}_\mathrm{QCD}^\mathrm{scat.}=\sum_{j=1}^{24}
\left(
C_{\mathrm{s},j}
\bar u B_j ^{(1)} u~
\bar v B_j ^{(2)} v
+
C_{\mathrm{o},j}
\bar u T^a B_j ^{(1)} u~
\bar v T^a B_j ^{(2)} v
\right)
\,,
\label{eq:m-scat}
\\
\mathcal{M}_\mathrm{QCD}^\mathrm{anni.}=\sum_{j=1}^{24}
\left(
C_{\mathrm{s},j}^c
\bar v B_j ^{(1)} u~
\bar u B_j ^{(2)} v
+
C_{\mathrm{o},j}^c
\bar v T^a B_j ^{(1)} u~
\bar u T^a B_j ^{(2)} v
\right)
\,.
\label{eq:m-anni}
\end{align}
where $u$ ($v$) is the quark (anti-quark) spinor and $2 T^a$ are the Gell-Mann
matrices.  The superscript ``c'' in Eq.~(\ref{eq:m-anni}) denotes that the
result is matched to the Lagrange density~(\ref{eq::Ldxy_2}), whereas in the
scattering channel we match our expressions to Eq.~(\ref{eq::Ldxy}).  The
coefficients $C_{\mathrm{s/o},j}$ and $C_{\mathrm{s/o},j}^c$, where ``s'' and
``o'' refer to singlet and octet colour states, are determined by an explicit
calculation of the amplitude in Eq.~(\ref{eq::q1q2-q1q2}).  
In calculating the QCD amplitude,
we treat the $\gamma$ matrices as $d$-dimensional objects
which satisfy 
\begin{align}
\{ \gamma^\mu, \gamma^\nu \} =2g^{\mu\nu},
\qquad g^\mu{}_\mu =d\,.
\end{align}
Unlike the case of 4-dimensional $\gamma$ matrices,
products of more than four 
$d$-dimensional $\gamma$ matrices
can not be expressed in terms of simpler products of $\gamma$ matrices,
and we have to treat all such products as independent basis elements.
Taking into account this fact,
we consider 
the following basis elements\footnote{Note that
  $B_{22}^{(1)}\otimes B_{22}^{(2)}$, $B_{23}^{(1)}\otimes B_{23}^{(2)}$ and
  $B_{24}^{(1)}\otimes B_{24}^{(2)}$ do not enter our calculation since, up to
  two-loop order, at most five $\gamma$ matrices are present in one fermion
  line. Nevertheless, for symmetry reasons, we provide also these basis
  elements.}
\begin{align}
B_1^{(1)}\otimes B_1^{(2)}
&=
1\!\!1
\otimes
1\!\!1
\,,\nonumber\\
B_2^{(1)}\otimes B_2^{(2)}
&=
{/\!\!\!v}
\otimes
1\!\!1
\,,\nonumber\\
B_3^{(1)}\otimes B_3^{(2)}
&=
1\!\!1
\otimes
{/\!\!\!v}
\,,\nonumber\\
B_4^{(1)}\otimes B_4^{(2)}
&=
{/\!\!\!v}
\otimes
{/\!\!\!v}
\,,\nonumber\\
B_5^{(1)}\otimes B_5^{(2)}
&=
\gamma^\mu
\otimes
\gamma_\mu
\,,\nonumber\\
B_6^{(1)}\otimes B_6^{(2)}
&=
\gamma^\mu
{/\!\!\!v}
\otimes
\gamma_\mu
\,,\nonumber\\
B_7^{(1)}\otimes B_7^{(2)}
&=
\gamma^\mu
\otimes
\gamma_\mu
{/\!\!\!v}
\,,\nonumber\\
B_8^{(1)}\otimes B_8^{(2)}
&=
\gamma^\mu
{/\!\!\!v}
\otimes
\gamma_\mu
{/\!\!\!v}
\,,\nonumber\\
B_9^{(1)}\otimes B_9^{(2)}
&=
\gamma^\mu\gamma^\nu
\otimes
\gamma_\mu\gamma_\nu
\,,\nonumber\\
B_{10}^{(1)}\otimes B_{10}^{(2)}
&=
\gamma^\mu\gamma^\nu
{/\!\!\!v}
\otimes
\gamma_\mu\gamma_\nu
\,,\nonumber\\
B_{11}^{(1)}\otimes B_{11}^{(2)}
&=
\gamma^\mu\gamma^\nu
\otimes
\gamma_\mu\gamma_\nu
{/\!\!\!v}
\,,\nonumber\\
B_{12}^{(1)}\otimes B_{12}^{(2)}
&=
\gamma^\mu\gamma^\nu
{/\!\!\!v}
\otimes
\gamma_\mu\gamma_\nu
{/\!\!\!v}
\,,\nonumber\\
B_{13}^{(1)}\otimes B_{13}^{(2)}
&=
\gamma^\mu\gamma^\nu\gamma^\rho
\otimes
\gamma_\mu\gamma_\nu\gamma_\rho
\,,\nonumber\\
B_{14}^{(1)}\otimes B_{14}^{(2)}
&=
\gamma^\mu\gamma^\nu\gamma^\rho
{/\!\!\!v}
\otimes
\gamma_\mu\gamma_\nu\gamma_\rho
\,,\nonumber\\
B_{15}^{(1)}\otimes B_{15}^{(2)}
&=
\gamma^\mu\gamma^\nu\gamma^\rho
\otimes
\gamma_\mu\gamma_\nu\gamma_\rho
{/\!\!\!v}
\,,\nonumber\\
B_{16}^{(1)}\otimes B_{16}^{(2)}
&=
\gamma^\mu\gamma^\nu\gamma^\rho
{/\!\!\!v}
\otimes
\gamma_\mu\gamma_\nu\gamma_\rho
{/\!\!\!v}
\,,\nonumber\\
B_{17}^{(1)}\otimes B_{17}^{(2)}
&=
\gamma^\mu\gamma^\nu\gamma^\rho\gamma^\sigma
\otimes
\gamma_\mu\gamma_\nu\gamma_\rho\gamma_\sigma
\,,\nonumber\\
B_{18}^{(1)}\otimes B_{18}^{(2)}
&=
\gamma^\mu\gamma^\nu\gamma^\rho\gamma^\sigma
{/\!\!\!v}
\otimes
\gamma_\mu\gamma_\nu\gamma_\rho\gamma_\sigma
\,,\nonumber\\
B_{19}^{(1)}\otimes B_{19}^{(2)}
&=
\gamma^\mu\gamma^\nu\gamma^\rho\gamma^\sigma
\otimes
\gamma_\mu\gamma_\nu\gamma_\rho\gamma_\sigma
{/\!\!\!v}
\,,\nonumber\\
B_{20}^{(1)}\otimes B_{20}^{(2)}
&=
\gamma^\mu\gamma^\nu\gamma^\rho\gamma^\sigma
{/\!\!\!v}
\otimes
\gamma_\mu\gamma_\nu\gamma_\rho\gamma_\sigma
{/\!\!\!v}
\,,\nonumber\\
B_{21}^{(1)}\otimes B_{21}^{(2)}
&=
\gamma^\mu\gamma^\nu\gamma^\rho\gamma^\sigma\gamma^\lambda
\otimes
\gamma_\mu\gamma_\nu\gamma_\rho\gamma_\sigma\gamma_\lambda
\,,\nonumber\\
B_{22}^{(1)}\otimes B_{22}^{(2)}
&=
\gamma^\mu\gamma^\nu\gamma^\rho\gamma^\sigma\gamma^\lambda
{/\!\!\!v}
\otimes
\gamma_\mu\gamma_\nu\gamma_\rho\gamma_\sigma\gamma_\lambda
\,,\nonumber\\
B_{23}^{(1)}\otimes B_{23}^{(2)}
&=
\gamma^\mu\gamma^\nu\gamma^\rho\gamma^\sigma\gamma^\lambda
\otimes
\gamma_\mu\gamma_\nu\gamma_\rho\gamma_\sigma\gamma_\lambda
{/\!\!\!v}
\,,\nonumber\\
B_{24}^{(1)}\otimes B_{24}^{(2)}
&=
\gamma^\mu\gamma^\nu\gamma^\rho\gamma^\sigma\gamma^\lambda
{/\!\!\!v}
\otimes
\gamma_\mu\gamma_\nu\gamma_\rho\gamma_\sigma\gamma_\lambda
{/\!\!\!v}
\,,
\end{align}
where 
${/\!\!\!v}={/\!\!\!p}/m$ and
the superscript refers to the fermion line.  We have explicitly
introduced the external momentum $p$ since we do not use the Dirac equation in
the course of the computation of the Feynman diagrams.

In matching to the NRQCD amplitude,
we use the following representation of the $\gamma$ matrices 
\begin{align}
\gamma^0=
\left(
\begin{array}{rr}
1&0\\
0&-1
\end{array}
\right)
,\quad
\vec \gamma=
\left(
\begin{array}{rr}
0&\vec \sigma\\
-\vec \sigma &0
\end{array}
\right)
\end{align}
in terms of
$(d-1)$-dimensional Pauli matrices
which satisfy 
\begin{align}
\{ \sigma^j, \sigma^k \} =2\delta^{jk},
\qquad \delta^{jj} =d-1\,.
\end{align}
In particular, we do not use
the commutation relation of the Pauli matrices
at this point.

The NRQCD amplitudes for the scattering and annihilation channels
can be written as
\begin{align}
&\mathcal{M}_\mathrm{NRQCD}^\mathrm{scat.}=
(\sqrt{2m})^4
\sum_{k=0}^{2}
\left(
c_{\mathrm{s},k}
\phi^\dagger \Sigma_k^{(1)} \phi~
\eta^\dagger \Sigma_k^{(2)} \eta~
+
c_{\mathrm{o},k}
\phi^\dagger T^a \Sigma_k^{(1)} \phi~
\eta^\dagger T^a \Sigma_k^{(2)} \eta~
\right)
\,,
\label{eq:mNR-scat}
\\
&\mathcal{M}_\mathrm{NRQCD}^\mathrm{anni.}=
(\sqrt{2m})^4
\sum_{k=0}^{2}
\left(
c_{\mathrm{s},k}^c
\eta^\dagger \Sigma_k^{c,(1)} \phi~
\phi^\dagger \Sigma_k^{c,(2)} \eta~
+
c_{\mathrm{o},k}^c
\eta^\dagger T^a \Sigma_k^{c,(1)} \phi~
\phi^\dagger T^a \Sigma_k^{c,(2)} \eta~
\right)
\,,
\label{eq:mNR-anni}
\end{align}
where $\phi$ and $\eta$ are two-component spinors which
in the limit of vanishing 3-momentum
are related to the $u$ and $v$ spinors in full QCD via
\begin{align}
u(p)=\sqrt{2m}
\left(
\begin{array}{c} \phi \\0 \end{array}
\right)
\,,\quad
v(p)=\sqrt{2m}
\left(
\begin{array}{c} 0\\ \eta \end{array}
\right)
\,.
\label{eq::uv2phieta}
\end{align}
The factor $\sqrt{2m}$ for each external quark appears due to our convention
for the normalization of the non-relativistic quark fields~\cite{Beneke:2013jia}.
Note that in Eqs.~(\ref{eq:mNR-scat}) and~(\ref{eq:mNR-anni}) different bases
have been introduced for the scattering and annihilation channels (see
also Eqs.~(\ref{eq::Ldxy}) and~(\ref{eq::Ldxy_2})).  In $d=4-2\epsilon$ dimensions
the basis elements are related to the Pauli matrices as
\begin{align}
&\Sigma_0^{(1)}\otimes \Sigma_0^{(2)}
=
1\!\!1
\otimes
1\!\!1
\,,\nonumber\\
&\Sigma_1^{(1)}\otimes \Sigma_1^{(2)}
=
-\frac{1}{8}
[\sigma^i,\sigma^j]
\otimes
[\sigma^i,\sigma^j]
\,,\nonumber\\
&\Sigma_2^{(1)}\otimes \Sigma_2^{(2)}
=
\frac{1}{64}
[\sigma^i,\sigma^j]
[\sigma^k,\sigma^l]
\otimes
[\sigma^i,\sigma^j]
[\sigma^k,\sigma^l]
\nonumber
\,,
\\
&\Sigma_0^{c,(1)}\otimes \Sigma_0^{c,(2)}
=
\sigma^i
\otimes
\sigma^i
\,,\nonumber\\
&\Sigma_1^{c,(1)}\otimes \Sigma_1^{c,(2)}
=
-\frac{1}{8}
[\sigma^i,\sigma^j]
\sigma^k
\otimes
[\sigma^i,\sigma^j]
\sigma^k
\,,\nonumber\\
&\Sigma_2^{c,(1)}\otimes \Sigma_2^{c,(2)}
=
\frac{1}{64}
[\sigma^i,\sigma^j]
[\sigma^k,\sigma^l]
\sigma^n
\otimes
[\sigma^i,\sigma^j]
[\sigma^k,\sigma^l]
\sigma^n
\,.
\label{eq::sigma_d}
\end{align}
For the two-loop calculation of $d_{xy}$ and $d^c_{xy}$ only $\Sigma_i$ and $\Sigma_i^{c}$
with $i=0,1,2$ are needed. At three loops basis elements constructed from
products of more than five Pauli matrices are necessary.

In order to obtain the matching coefficients in Eqs.~\eqref{eq::Ldxy}
and~\eqref{eq::Ldxy_2}, one has to reduce the structure of the Pauli matrices
to $1\!\!1\otimes 1\!\!1$ and $\sigma^j\otimes\sigma^j$ instead of those in
Eqs.~\eqref{eq::sigma_d}.  In other words, one has to take the limit $d\to 4$.
There are different prescriptions to do this; one can use the
commutation relation $[\sigma_j,\sigma_k]=2i\varepsilon_{jkl}\sigma_l$
assuming
$\varepsilon^{jkl}\varepsilon^{jkl'}=(d-2)\delta^{ll'}$~\cite{Pineda:1998kj},
or $\varepsilon^{jkl}\varepsilon^{jkl'}=2\delta^{ll'}$.  Since it is unclear
which prescription should be used, we provide the $d$-dimensional
results in the basis of Eqs.~\eqref{eq::sigma_d}.  Nevertheless, it is useful to have
the conventional matching coefficients $d_{xy}$.  For this purpose we
adopt $\varepsilon^{jkl}\varepsilon^{jkl'}=2\delta^{ll'}$ and obtain
\begin{align}
&\Sigma_1^{(1)}\otimes \Sigma_1^{(2)}
=
\sigma^j
\otimes
\sigma^j
\,,\nonumber\\
&\Sigma_2^{(1)}\otimes \Sigma_2^{(2)}
=
31\!\!1
\otimes
1\!\!1
-2
\sigma^j
\otimes
\sigma^j
\nonumber
\,,
\\
&\Sigma_1^{c,(1)}\otimes \Sigma_1^{c,(2)}
=
31\!\!1
\otimes
1\!\!1
-2
\sigma^j
\otimes
\sigma^j
\,,\nonumber
\\
&\Sigma_2^{c,(1)}\otimes \Sigma_2^{c,(2)}
=
-61\!\!1
\otimes
1\!\!1
+7
\sigma^j
\otimes
\sigma^j
\,.
\label{eq::sigma_4}
\end{align}
In the following, 
we refer to this prescription
as ``taking the limit $d\to 4$".

At this point it is convenient to discuss the scattering and
annihilation channel separately. In the former case one has to
consider $\gamma^{\mu_1}\cdots\gamma^{\mu_n}$ sandwiched between $\bar
u$ and $u$ or $\bar v$ and $v$, which means that only diagonal parts
of $\gamma^{\mu_1}\cdots\gamma^{\mu_n}$ contribute. 
Then we obtain
\begin{align}
\bar u(p) B_j^{(1)}u(p)
~\bar v(p) B_j^{(2)} v(p)
=
\sum_{k=0}^2
R_j^k~
\phi^\dagger \Sigma_k^{(1)}\phi~ 
\eta^\dagger \Sigma_k^{(2)}\eta\,,
\label{eq:basis-change-scat}
\end{align}
where 
the $R_j^k$ are given in Tab.~\ref{tab:rjk-scat}.
In order to obtain the table entries one can use
the equation of motion for the external fermions
\begin{align}
/\!\!\! v u(p)=u(p)
\,,\quad 
/\!\!\! v v(p)=-v(p)
\,.
\end{align}
Afterwards, we
insert the explicit expressions for the spinors $u$ and $v$ in terms of
$\phi$ and $\eta$ (cf. Eq.~(\ref{eq::uv2phieta})).
After substituting Eq.~\eqref{eq:basis-change-scat} 
into Eq.~\eqref{eq:m-scat}
and comparing with Eq.~\eqref{eq:mNR-scat},
we obtain the relations between 
NRQCD coefficients $c_{\mathrm{s/o},k}$
and QCD coefficients $C_{\mathrm{s/o},j}$:
\begin{align}
c_{\mathrm{s/o},k}
=\sum_{j=1}^{24}
R_j^k~C_{\mathrm{s/o},j}
\,.
\label{eq::match}
\end{align}


\begin{table}[t]
\centering
\begin{tabular}{c}
\begin{minipage}{0.15\hsize}
\centering
\begin{tabular}{c|c||c}
\multicolumn{2}{c||}{} & $k$ \\\cline{3-3}
\multicolumn{2}{c||}{} & 0 \\\hline \hline
\multirow{8}{*}{$j$}
& \multirow{1}{*}{1} 
&$-1$\\
& \multirow{1}{*}{2} 
&$-1$\\
& \multirow{1}{*}{3} 
&$1$\\
& \multirow{1}{*}{4} 
&$1$\\
& \multirow{1}{*}{5} 
&$1$\\
& \multirow{1}{*}{6} 
&$1$\\
& \multirow{1}{*}{7} 
&$-1$\\
& \multirow{1}{*}{8} 
&$-1$\\
\end{tabular}
\end{minipage}
\begin{minipage}{0.3\hsize}
\centering
\begin{tabular}{r||rr}
& \multicolumn{2}{c}{$k$} \\\cline{2-3}
& 0 & 1 \\\hline \hline
9&$-d$&$2$\\
10&$-d$&$2$\\
11&$d$&$-2$\\
12&$d$&$-2$\\
13&$3d-2$&$-6$\\
14&$3d-2$&$-6$\\
15&$-3d+2$&$6$\\
16&$-3d+2$&$6$
\end{tabular}
\end{minipage}
\begin{minipage}{0.5\hsize}
\centering
\begin{tabular}{c||rrr}
& \multicolumn{3}{c}{$k$} \\\cline{2-4}
& 0 & 1 & 2\\\hline \hline
17&$-d^2-4d+4$&$4d+8$&$-4$\\
18&$-d^2-4d+4$&$4d+8$&$-4$\\
19&$d^2+4d-4$&$-4d-8$&$4$\\
20&$d^2+4d-4$&$-4d-8$&$4$\\
21&$5d^2-4$&$-20d$&$20$\\
22&$5d^2-4$&$-20d$&$20$\\
23&$-5d^2+4$&$20d$&$-20$\\
24&$-5d^2+4$&$20d$&$-20$
\end{tabular}
\end{minipage}
\end{tabular}
\caption{The coefficients $R_j^k$ introduced in
  Eq.~(\ref{eq:basis-change-scat}) for the matching of the scattering amplitude.}
\label{tab:rjk-scat}
\end{table}


In the case of the annihilation channel
$\gamma^{\mu_1}\cdots\gamma^{\mu_n}$ is sandwiched between $\bar v$
and $u$ or $\bar u$ and $v$ and thus only the off-diagonal
parts contribute, which means that one needs an odd number of $\vec
\gamma$ matrices. In analogy to Eq.~(\ref{eq:basis-change-scat})
we can write
\begin{align}
\bar v(p)B_j^{(1)}u(p)~
\bar u(p) B_j^{(2)}v(p)
=
\sum_{k=0}^2
R_j^{c,k}~
\eta^\dagger \Sigma_k^{c,(1)}\phi~
\phi^\dagger \Sigma_k^{c,(2)}\eta\,,
\label{eq:basis-change-anni}
\end{align}
where $R_j^{c,k}$ are given in Tab.~\ref{tab:rjk-anni}.
Substituting Eq.~\eqref{eq:basis-change-anni} into
Eq.~\eqref{eq:m-anni} and comparing with Eq.~\eqref{eq:mNR-anni} leads
to the relations between NRQCD coefficients
$c^c_{\mathrm{s/o},k}$ and QCD coefficients
$C^c_{\mathrm{s/o},j}$:
\begin{align}
c^c_{\mathrm{s/o},k}
=\sum_{j=1}^{24}
R_j^{c,k}~C^c_{\mathrm{s/o},j}
\,.
\label{eq::match_c}
\end{align}
Results up to two loops for $c_{\mathrm{s/o},k}$ and $c^c_{\mathrm{s/o},k}$ are presented in
Section~\ref{sec::res}.



\begin{table}[t]
\centering
\begin{tabular}{c}
\begin{minipage}{0.18\hsize}
\centering
\begin{tabular}{c|c||c}
\multicolumn{2}{c||}{} & $k$ \\\cline{3-3}
\multicolumn{2}{c||}{} & 0 \\\hline \hline
\multirow{8}{*}{$j$}
& \multirow{1}{*}{1} 
&\\
& \multirow{1}{*}{2} 
&\\
& \multirow{1}{*}{3} 
&\\
& \multirow{1}{*}{4} 
&\\
& \multirow{1}{*}{5} 
&$-1$\\
& \multirow{1}{*}{6} 
&$-1$\\
& \multirow{1}{*}{7} 
&$1$\\
& \multirow{1}{*}{8} 
&$1$\\
\end{tabular}
\end{minipage}
\begin{minipage}{0.3\hsize}
\centering
\begin{tabular}{r||rr}
& \multicolumn{2}{c}{$k$} \\\cline{2-3}
& 0 & 1 \\\hline \hline
9&$2$&\\
10&$2$&\\
11&$-2$&\\
12&$-2$&\\
13&$-d-2$&$2$\\
14&$-d-2$&$2$\\
15&$d+2$&$-2$\\
16&$d+2$&$-2$\\
\end{tabular}
\end{minipage}
\begin{minipage}{0.45\hsize}
\centering
\begin{tabular}{c||rrr}
& \multicolumn{3}{c}{$k$} \\\cline{2-4}
& 0 & 1 & 2\\\hline \hline
17&$4d$&$-8$\\
18&$4d$&$-8$\\
19&$-4d$&$8$\\
20&$-4d$&$8$\\
21&$-d^2-8d+4$&$4d+16$&$-4$\\
22&$-d^2-8d+4$&$4d+16$&$-4$\\
23&$d^2+8d-4$&$-4d-16$&$4$\\
24&$d^2+8d-4$&$-4d-16$&$4$\\
\end{tabular}
\end{minipage}
\end{tabular}
\caption{The coefficients $R_j^{c,k}$ introduced in
  Eq.~(\ref{eq:basis-change-anni}) for the matching of the annihilation amplitude.}
\label{tab:rjk-anni}
\end{table}



\subsection{Loop integrals}

In the following we briefly describe the workflow of our calculation.  We
first generate the full QCD amplitudes with {\tt
  qgraf}~\cite{Nogueira:1991ex} and map the output to general four-point
families which have four and nine independent propagators at one and two
loops, respectively.  Next, we apply projectors to obtain the coefficients of
the basis elements $B_i$ which leads us to scalar expressions.  Afterwards, we
specify the kinematics given in Eq.~(\ref{eq::q1q2-q1q2}). At two loops this
leads to five (instead of nine) linearly independent propagators.  One has to
apply a partial fraction decomposition in order to obtain integral families
which can be reduced to master integrals using {\tt
  FIRE}~\cite{Smirnov:2014hma} and \texttt{LiteRed}~\cite{Lee:2013mka}.

In an alternative approach, which we use for some of the integral families, we
specify only some of the kinematic relations such that the propagators are
still linearly independent. Then we perform an integration-by-parts
reduction, apply the full kinematic information of
Eq.~(\ref{eq::q1q2-q1q2}) to the resulting master integrals, perform a partial
fraction decomposition to these masters, and a further (very simple) reduction in order to
arrive at the same set of master integrals as in our standard approach.  Note
that in all cases the reduction problem is quite simple and takes at most, even for
general QCD gauge parameter, a few minutes on a desktop computer.

\begin{figure}[t]
  \centering
  \includegraphics[width=\textwidth]{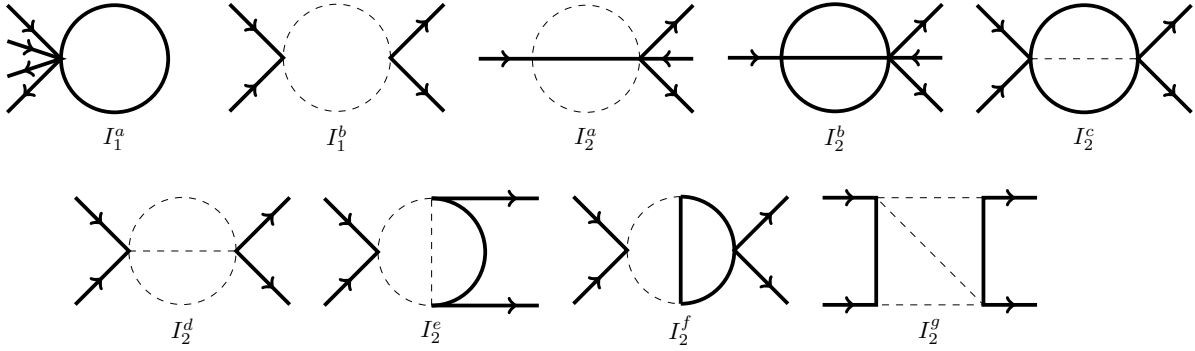}
  \caption{\label{fig::masters} One- and two-loop irreducible master
    integrals.  At two-loop order, there are also three reducible master
    integrals: $(I_1^a)^2, I_1^aI_1^b, (I_1^b)^2$.  Solid and dashed lines
    represent massive and massless lines, respectively.  Each external line carries the
    momentum $p$.  For the scattering channel only
    $I_1^a, I_2^a, I_2^b, I_2^c$ are needed, and in the annihilation
    contribution all master integrals appear.}
\end{figure}

Our final result for the QCD amplitude can be expressed in terms of two
one-loop and ten two-loop master integrals (cf. Fig.~\ref{fig::masters}).
We retain the exact $\epsilon$-dependence up to this point 
and provide the corresponding results in an ancillary file~\cite{progdata}.
Most of the master integrals are available in the
literature~\cite{Czarnecki:1997vz,Beneke:1997jm,Piclum:2007an}.  
However, not all of them are known analytically, and for some higher orders in $\epsilon$ are needed.
Furthermore, to our knowledge the
box-type integral $I_2^g$ is not available in the literature so far.  For this
reason we (re)compute those integrals analytically and present the results in
Appendix~\ref{app::master}.

After inserting the master integrals into the four-fermion amplitudes we use
Eqs.~(\ref{eq::match}) and~(\ref{eq::match_c}), expand
in $\epsilon$ and thus obtain the matching coefficients $c_{s/o,k}$ and
$c^c_{s/o,k}$. Analytic results are presented in Section~\ref{sec::res}.
Let us mention that the colour and Lorentz part of the QCD amplitude factorizes
such that they can be computed independently.



\section{\label{sec::cD}Gluon fermion matching coefficients}

The purpose of this section is the computation of $c_D$
which has to be combined with $d_{vs}$ in order to cancel the $\xi$
dependence. Since the calculation of $c_F$ and $c_S$ proceeds among similar
lines we compute all three matching coefficients simultaneously
and present results up to two loops.

The matching coefficients $c_D$, $c_F$ and $c_S$ can be extracted from
the gluon-quark vertex function which we parameterize as
\begin{eqnarray}
  \Gamma^a_\mu 
  &=& ig_s 
      \bar{u}(p^{\prime}) T^a \left[\gamma^{\mu}F_1\left(\frac{q^2}{m^2}\right)
      +\frac{i\sigma^{\mu\nu}q_{\nu}}{2m}F_2\left(\frac{q^2}{m^2}\right)\right]u(p),
      \label{eq::Gamma}
\end{eqnarray}
where $p$ ($p^\prime$) is the outgoing (incoming) quark (anti-quark) momentum
and $q=p-p^\prime$. The quark momenta are on-shell,
i.e. $p^2=(p^\prime)^2=m^2$ and we have
$\sigma^{\mu\nu} = i[\gamma^\mu,\gamma^\nu]/2$.
The fundamental indices in the matrix $T^a$ are suppressed.

The calculation is performed in the background field
method~\cite{Abbott:1980hw} where the gauge parameter $\xi$ enters via the
gluon propagator
\begin{align}
D_{g}^{\mu \nu}(q)=\frac{-i}{q^{2}+i \varepsilon}
\left( g^{\mu \nu}-\xi \frac{q^{\mu} q^{\nu}}{q^{2}} \right)
\end{align}
and the vertex of the background gluon and two quantum gluons, which contains
a factor $1/(1-\xi)$.  Note that the $\xi$-dependence is treated exactly throughout
the calculation.

For the matching calculation it is sufficient to consider $\Gamma^a_\mu$ in
the limit of small gluon momentum $q$. In fact, after considering the
non-relativistic limit in Eq.~(\ref{eq::Gamma}) the comparison to the
tree-level Feynman rules from ${\cal L}_{\psi}$ in Eq.~(\ref{eq::LgQ})
leads to
\begin{eqnarray}
  \tilde{c}_F&=& 1+F_2(0) \,,\nonumber\\
  \tilde{c}_D&=& 1+2F_2(0)+8F_1'(0) -\frac{16d_2}{d_1}\,,\nonumber\\
  \tilde{c}_S&=& 1+2F_2(0)\,,
  \label{eq::cD}
\end{eqnarray}
where the prime indicates the derivative w.r.t.\ the argument and $d_1, d_2$ can be
found in Appendix~\ref{app::d_i}.
The tilde in Eq.~(\ref{eq::cD}) indicates that no rescaling
of the gluon field has been performed. Thus, in order to
obtain the matching coefficients present in the Lagrange density~(\ref{eq::LgQ})
one has to apply Eq.~(\ref{eq::A_rescale}) in Appendix~\ref{app::d_i}.
Note that $d_1=1+\mathcal{O}(\alpha_s)$ and $d_2=\mathcal{O}(\alpha_s)$,
and thus $d_2/d_1\to d_2$ at one-loop order.
We can Taylor-expand the form factors $F_1$ and $F_2$ in the gluon
momentum and are left with one- and two-loop on-shell integrals
which are well studied in the literature (see, e.g.,
Refs.~\cite{Gray:1990yh,Fleischer:1991xp}). 

In the following we provide results for the form factors and
their derivatives for $q^2=0$.
We parametrize the form factors as
\begin{align}
  F_{i}=\sum _{j\geq 1} \left( \frac{\alpha_s^{(n_l+n_h)} (m)}{\pi} \right) ^j 
    \left(\frac{\mu^2}{m^2}\right)^{j{\epsilon}}
F_{i}^{(j)}
  \,.
\end{align}
Note that the $F_{i}$ still contain poles and also have
an explicit $\mu$ dependence.
Below we show the $\epsilon$-expanded expressions
and provide 
the $\epsilon$-exact results in 
an ancillary file~\cite{progdata}.
Our results for $F_i^\prime(0)$ and $F_2(0)$ read
\begin{align}
F_1'^{(1)}(0)&=
C_A \left[-\frac{5}{48 \epsilon}-\frac{1}{16}\right]+C_F \left[-\frac{1}{6 \epsilon}-\frac{1}{8}\right]
+\mathcal{O}(\epsilon)
\,,\nonumber\\
F_2^{(1)}(0)&=
C_A \left[\frac{1}{4 \epsilon}+\frac{1}{2}\right]+\frac{C_F}{2}
+\mathcal{O}(\epsilon)
\,,\nonumber\\
F_1'^{(2)}(0)&=
C_F^2\left[-\frac{3 \zeta_3}{4}-\frac{47}{576}-\frac{175 \pi ^2}{864}+\frac{1}{2} \pi ^2 \log 2\right]
+C_F n_h T_F\left[\frac{l_\mu^2}{36}+\frac{3 \pi ^2}{32}-\frac{1099}{1296}\right]
\nonumber\\
&
+C_A C_F\left[\frac{1}{16 \epsilon^2}+\frac{\frac{\pi ^2}{72}-\frac{13}{48}}{\epsilon}-\frac{11}{144} l_\mu^2+\frac{29 \zeta_3}{48}+\frac{19 \pi ^2}{864}-\frac{1783}{5184}-\frac{7}{24} \pi ^2 \log 2\right]
\nonumber\\
&
+C_A^2\left[\frac{3}{128 \epsilon^2}+\frac{-\frac{71}{576}-\frac{\pi ^2}{576}}{\epsilon}-\frac{55 l_\mu^2}{1152}-\frac{5 \zeta_3}{96}+\frac{5 \pi ^2}{3456}-\frac{397}{324}+\frac{1}{48} \pi ^2 \log 2\right]
\nonumber\\
&
+C_A n_h T_F\left[-\frac{1}{720 \epsilon}+\frac{5 l_\mu^2}{288}-\frac{\pi ^2}{108}+\frac{2779}{16200}\right]
+\xi\left[C_A \left(\frac{1}{80 \epsilon}-\frac{13}{600}\right) n_h T_F-\frac{3 C_A^2}{256}\right]
\nonumber\\
&
+C_F n_l T_F\left[-\frac{1}{36 \epsilon^2}+\frac{5}{108 \epsilon}+\frac{l_\mu^2}{36}+\frac{\pi ^2}{54}+\frac{283}{1296}\right]
\nonumber\\
&
+C_A n_l T_F\left[-\frac{5}{288 \epsilon^2}+\frac{103}{864 \epsilon}+\frac{5 l_\mu^2}{288}+\frac{5 \pi ^2}{432}+\frac{1357}{5184}\right]
+\mathcal{O}(\epsilon)
\,,\nonumber\\
F_2^{(2)}(0)&=
C_F^2\left[\frac{3 \zeta_3}{4}-\frac{31}{16}+\frac{5 \pi ^2}{12}-\frac{1}{2} \pi ^2 \log 2\right]
+C_A C_F\left[\frac{1}{8 \epsilon}-\frac{\zeta_3}{8}+\frac{\pi ^2}{12}+\frac{341}{144}+\frac{1}{12} \pi ^2 \log 2\right]
\nonumber\\
&
+C_A^2\left[-\frac{1}{12 \epsilon^2}+\frac{35}{144 \epsilon}+\frac{11 l_\mu^2}{96}-\frac{\zeta_3}{8}-\frac{65 \pi ^2}{576}+\frac{859}{432}+\frac{1}{12} \pi ^2 \log 2\right]
-\frac{25}{36} C_F n_l T_F
\nonumber\\
&
+C_F n_h T_F\left[\frac{119}{36}-\frac{\pi ^2}{3}\right]
+C_A n_h T_F\left[-\frac{1}{24} l_\mu^2+\frac{\pi ^2}{16}-\frac{149}{216}\right]
\nonumber\\
&
+C_A n_l T_F\left[\frac{1}{24 \epsilon^2}-\frac{13}{144 \epsilon}-\frac{1}{24} l_\mu^2-\frac{\pi ^2}{36}-\frac{299}{432}\right]
+\mathcal{O}(\epsilon)
\,.\label{eq:f1pf2-2}
\end{align}

Our two-loop result for $F_2(0)$ agrees with Refs.~\cite{Czarnecki:1997dz,Grozin:2007fh} and
the QED part\footnote{The QED result is obtained for $C_A=0$, $C_F=1$,
  $T_F=1$, $n_l=0$, $n_h=1$ and the coupling constant renormalized in the on-shell scheme.}  
  of $F^\prime_1(0)$ can be found
in~\cite{Barbieri:1970vj,Melnikov:1999xp}. The two-loop QCD corrections to
$F^\prime_1(0)$ are new.

We can now use Eq.~(\ref{eq::cD}), apply the rescaling of Eq.~(\ref{eq::A_rescale})
and decouple the heavy quark
in the gluon wave function and the coupling constant\footnote{Note that we
  apply the decoupling also to the factor $g_s$ in Eq.~(\ref{eq::LgQ}).}
in order to compute $c_D$, $c_F$ and $c_S$.
In the following we present one- and two-loop expressions for
$c_D$ and postpone $c_F$ and $c_S$ to Appendix~\ref{app::cFcS}.
By parameterizing the matching coefficients $c_X$ as
\begin{align}
  c_{X}=1+\sum _{j\geq 1} \left( \frac{\alpha_s^{(n_l)} (m)}{\pi} \right) ^j 
    \left(\frac{\mu^2}{m^2}\right)^{j{\epsilon}}
  c_{X}^{(j)}
  \,,
\end{align}
we obtain for $c_D$
\begin{align}
c_D^{(1)}=&
C_A \left[\frac{1}{2}-\frac{1}{3 \epsilon}\right]-\frac{4 C_F}{3 \epsilon}-\frac{4 n_h T_F}{15}
+\mathcal{O}(\epsilon)
\,,\nonumber\\
c_D^{(2)}=&
C_F^2\left[-\frac{9 \zeta_3}{2}-\frac{163}{36}-\frac{85 \pi ^2}{108}+3 \pi ^2 \log 2\right]
+C_F n_l T_F\left[-\frac{2}{9 \epsilon^2}+\frac{10}{27 \epsilon}+\frac{2 l_\mu^2}{9}+\frac{4 \pi ^2}{27}+\frac{29}{81}\right]
\nonumber\\
&
+C_A C_F\left[\frac{1}{2 \epsilon^2}+\frac{\frac{\pi ^2}{9}-\frac{23}{12}}{\epsilon}-\frac{11}{18} l_\mu^2+\frac{55 \zeta_3}{12}+\frac{37 \pi ^2}{108}+\frac{643}{324}-\frac{13}{6} \pi ^2 \log 2\right]
\nonumber\\
&
+C_A^2\left[\frac{1}{48 \epsilon^2}+\frac{-\frac{1}{2}-\frac{\pi ^2}{72}}{\epsilon}-\frac{11}{72} l_\mu^2-\frac{2 \zeta_3}{3}-\frac{185 \pi ^2}{864}-\frac{3775}{648}+\frac{1}{3} \pi ^2 \log 2\right]
-\frac{4 n_h n_l T_F^2}{45 \epsilon}
\nonumber\\
&
+C_A n_l T_F\left[-\frac{1}{18 \epsilon^2}+\frac{167}{216 \epsilon}+\frac{l_\mu^2}{18}+\frac{\pi ^2}{27}+\frac{115}{162}\right]
+\xi\left[C_A \left(\frac{1}{20 \epsilon}-\frac{13}{150}\right) n_h T_F-\frac{3 C_A^2}{32}\right]
\nonumber\\
&
+C_F n_h T_F\left[\frac{5 \pi ^2}{108}-\frac{32}{27}\right]
+C_A n_h T_F\left[-\frac{1}{36 \epsilon}+\frac{\pi ^2}{24}+\frac{1613}{5400}\right]
+\mathcal{O}(\epsilon)
\,.\label{eq:cD-2}
\end{align}
Note the $\xi$ dependence in the second last line which is inherited from
$F_1'^{(2)}(0)$ and $d_2$ according to Eq.~(\ref{eq::cD}).


\section{\label{sec::res}Results for the four-fermion matching coefficients}

In this section we present first our results in $d$ dimensions
and afterwards take the limit $d\to4$. We discuss both the scattering and the
annihilation channel.

\subsection{\label{sub::c_xy}NRQCD four quark coefficients in $d$ dimensions}

We parametrize the matching coefficients as follows
\begin{align}
  c_{\mathrm{s/o},k} =\sum _{j\geq 0} \pi^2 \left( \frac{\alpha_s^{(n_l)}(m )}{\pi} \right) ^{j+1}
    \left(\frac{\mu^2}{m^2}\right)^{j{\epsilon}}
  c_{\mathrm{s/o},k}^{(j)}
  \,,
\end{align}
and use an analogous expansion for $c_{\mathrm{s/o},k}^c$.  
At tree level we have
\begin{align}
  c_{\mathrm{o},0}^{c,(0)} = -1\,,
\end{align}
and all the other coefficients are zero.  We have obtained exact results in
$d$ dimensions both at one and two loops and provide the corresponding results
in an ancillary file~\cite{progdata}.  Below we show the
$\epsilon$-expanded expressions.

\subsubsection{One-loop results}

Our one-loop results for the scattering channel are given by
\begin{align}
c_{s,0}^{(1)}=&
\frac{C_F}{N_c} \left[\frac{1}{2 \epsilon}+\frac{1}{3}\right]
+\mathcal{O}(\epsilon)
\,,\nonumber\\
c_{s,1}^{(1)}=&
\frac{C_F}{2N_c}
+\mathcal{O}(\epsilon)
\,,\nonumber\\
c_{o,0}^{(1)}=&
C_A \left[\frac{11}{12}-\frac{5}{4 \epsilon}\right]+C_F \left[\frac{2}{\epsilon}+\frac{4}{3}\right]
+\mathcal{O}(\epsilon)
\,,\nonumber\\
c_{o,1}^{(1)}=&
C_A \left[-\frac{1}{4 \epsilon}-\frac{1}{2}\right]+2 C_F
+\mathcal{O}(\epsilon)
\,.\label{eq:c-1}
\end{align}

Note that $c_{s,2}^{(1)}=0$ and $c_{o,2}^{(1)}=0$ since at one-loop
order at most two $\sigma$ matrices are present in a spinor line.  
In the literature,
the factor $1/N_c$ in the colour singlet matching coefficients 
are expressed as $(C_A-2C_F)$.
Here and in the following, 
we use  $1/N_c$ in order to have more compact expressions.

For the annihilation channel we have
\begin{align}
c_{s,0}^{c,(1)}=&
\frac{C_F}{N_c} \left[\frac{2}{3}+\frac{i \pi }{3}-\frac{2 \log 2}{3}\right]
+\mathcal{O}(\epsilon)
\,,\nonumber\\
c_{s,1}^{c,(1)}=&
\frac{C_F}{N_c} \left[\frac{1}{3}+\frac{i \pi }{6}-\frac{\log 2}{3}\right]
+\mathcal{O}(\epsilon)
\,,\nonumber\\
c_{o,0}^{c,(1)}=&
C_A \left[-\frac{145}{36}-\frac{i \pi }{2}+\log 2\right]+C_F \left[\frac{20}{3}+\frac{4 i \pi }{3}-\frac{8 \log 2}{3}\right]
\nonumber\\
&\qquad +\frac{8 n_h T_F}{9}+n_l T_F \left[\frac{5}{9}+\frac{i \pi }{3}-\frac{2 \log 2}{3}\right]
+\mathcal{O}(\epsilon)
\,,\nonumber\\
c_{o,1}^{c,(1)}=&
C_A \left[-\frac{1}{2}-\frac{i \pi }{4}+\frac{\log 2}{2}\right]+C_F \left[\frac{4}{3}+\frac{2 i \pi }{3}-\frac{4 \log 2}{3}\right]
+\mathcal{O}(\epsilon)
\,.\label{eq:cc-1}
\end{align}

where we have again $c_{s,2}^{c,(1)}=0$ and $c_{o,2}^{c,(1)}=0$.

\subsubsection{Two-loop results}

At two-loop order the matching coefficients obtained form the scattering
process read
\begin{align}
c_{s,0}^{(2)}=&
\frac{C_F^2}{N_c}\left[-\frac{3 \pi ^2}{16 \epsilon}+\frac{33 \zeta_3}{16}+\frac{23 \pi ^2}{48}-\frac{63}{4}+\frac{21}{8} \pi ^2 \log 2\right]
+\frac{C_F n_h T_F}{N_c}\left[\frac{\pi ^2}{9}-\frac{20}{27}\right]
\nonumber\\
&
+\frac{C_A C_F}{N_c}\left[-\frac{11}{24 \epsilon^2}+\frac{-\frac{8}{9}-\frac{47 \pi ^2}{192}}{\epsilon}+\frac{11 l_\mu^2}{24}-\frac{503 \zeta_3}{64}+\frac{1739 \pi ^2}{576}+\frac{809}{24}-\frac{19}{32} \pi ^2 \log 2\right]
\nonumber\\
&
+\frac{C_F n_l T_F}{N_c}\left[\frac{1}{6 \epsilon^2}-\frac{7}{18 \epsilon}-\frac{1}{6} l_\mu^2-\frac{\pi ^2}{9}-\frac{19}{9}\right]
+\mathcal{O}(\epsilon)
\,,\nonumber\\
c_{s,1}^{(2)}=&
\frac{C_F^2}{N_c}\left[\frac{5 \pi ^2}{24 \epsilon}+\frac{27 \zeta_3}{8}+\frac{45 \pi ^2}{16}-\frac{5}{12}-\frac{31}{12} \pi ^2 \log 2\right]
-\frac{5 C_F n_h T_F}{9 N_c}
+\frac{4 C_F n_l T_F}{9 N_c}
\nonumber\\
&
+\frac{C_A C_F}{N_c}\left[\frac{11 \pi ^2}{96 \epsilon}+\frac{89 \zeta_3}{32}-\frac{29 \pi ^2}{72}-\frac{17}{36}+\frac{55}{48} \pi ^2 \log 2\right]
+\mathcal{O}(\epsilon)
\,,\nonumber\\
c_{s,2}^{(2)}=&
\frac{C_F^2}{N_c}\left[\frac{\pi ^2}{16 \epsilon}-\frac{3 \zeta_3}{16}+\frac{29 \pi ^2}{48}+\frac{1}{2}-\frac{7}{8} \pi ^2 \log 2\right]
\nonumber\\
&
+\frac{C_A C_F}{N_c}\left[-\frac{\pi ^2}{64 \epsilon}+\frac{9 \zeta_3}{64}-\frac{41 \pi ^2}{192}-\frac{1}{4}+\frac{13}{32} \pi ^2 \log 2\right]
+\mathcal{O}(\epsilon)
\,,\nonumber\\
c_{o,0}^{(2)}=&
C_F^2\left[-\frac{9 \pi ^2}{16 \epsilon}+\frac{171 \zeta_3}{16}+\frac{193 \pi ^2}{48}-56+\frac{63}{8} \pi ^2 \log 2\right]
+C_A n_h T_F\left[-\frac{1}{5 \epsilon}-\frac{5 \pi ^2}{18}+\frac{1289}{675}\right]
\nonumber\\
&
+C_A C_F\left[-\frac{7}{3 \epsilon^2}+\frac{-\frac{97}{18}-\frac{53 \pi ^2}{96}}{\epsilon}+\frac{11 l_\mu^2}{6}-\frac{1211 \zeta_3}{32}+\frac{2293 \pi ^2}{288}+\frac{2683}{18}-\frac{127}{16} \pi ^2 \log 2\right]
\nonumber\\
&
+C_A^2\left[\frac{49}{48 \epsilon^2}+\frac{\frac{7}{18}+\frac{13 \pi ^2}{64}}{\epsilon}-\frac{55}{48} l_\mu^2+\frac{633 \zeta_3}{64}-\frac{1505 \pi ^2}{576}-\frac{3269}{72}+\frac{37}{32} \pi ^2 \log 2\right]
\nonumber\\
&
+C_F n_l T_F\left[\frac{2}{3 \epsilon^2}-\frac{14}{9 \epsilon}-\frac{2}{3} l_\mu^2-\frac{4 \pi ^2}{9}-\frac{76}{9}\right]
+\xi\left[\frac{3 C_A^2}{32}+C_A \left(\frac{13}{150}-\frac{1}{20 \epsilon}\right) n_h T_F\right]
\nonumber\\
&
+C_F n_h T_F\left[\frac{4 \pi ^2}{9}-\frac{80}{27}\right]
+C_A n_l T_F\left[-\frac{5}{12 \epsilon^2}+\frac{35}{36 \epsilon}+\frac{5 l_\mu^2}{12}+\frac{5 \pi ^2}{18}+\frac{77}{18}\right]
+\mathcal{O}(\epsilon)
\,,\nonumber\\
c_{o,1}^{(2)}=&
C_F^2\left[\frac{5 \pi ^2}{8 \epsilon}+\frac{77 \zeta_3}{8}+\frac{121 \pi ^2}{12}-\frac{11}{6}-\frac{109}{12} \pi ^2 \log 2\right]
+C_A n_h T_F\left[\frac{35}{27}-\frac{\pi ^2}{18}\right]
\nonumber\\
&
+C_A C_F\left[\frac{\frac{3 \pi ^2}{16}-\frac{1}{4}}{\epsilon}+\frac{83 \zeta_3}{16}-\frac{929 \pi ^2}{144}-\frac{5}{36}+\frac{229}{24} \pi ^2 \log 2\right]
\nonumber\\
&
+C_A^2\left[\frac{1}{6 \epsilon^2}+\frac{-\frac{11}{72}-\frac{25 \pi ^2}{288}}{\epsilon}-\frac{11}{48} l_\mu^2-\frac{139 \zeta_3}{96}+\frac{955 \pi ^2}{864}-\frac{5}{108}-\frac{103}{48} \pi ^2 \log 2\right]
\nonumber\\
&
-\frac{20}{9} C_F n_h T_F
+\frac{16 C_F n_l T_F}{9}
+C_A n_l T_F\left[-\frac{1}{12 \epsilon^2}+\frac{1}{18 \epsilon}+\frac{l_\mu^2}{12}+\frac{\pi ^2}{18}-\frac{31}{54}\right]
+\mathcal{O}(\epsilon)
\,,\nonumber\\
c_{o,2}^{(2)}=&
C_F^2\left[\frac{3 \pi ^2}{16 \epsilon}-\frac{9 \zeta_3}{16}+\frac{29 \pi ^2}{16}+\frac{3}{2}-\frac{21}{8} \pi ^2 \log 2\right]
\nonumber\\
&
+C_A C_F\left[-\frac{3 \pi ^2}{32 \epsilon}+\frac{57 \zeta_3}{32}-\frac{127 \pi ^2}{96}-\frac{7}{4}+\frac{37}{16} \pi ^2 \log 2\right]
\nonumber\\
&
+C_A^2\left[\frac{\pi ^2}{64 \epsilon}-\frac{27 \zeta_3}{64}+\frac{15 \pi ^2}{64}+\frac{1}{2}-\frac{15}{32} \pi ^2 \log 2\right]
+\mathcal{O}(\epsilon)
\,.\label{eq:c-2}
\end{align}

\begin{figure}[t]
  \centering
  \begin{tabular}{ccc}
    \includegraphics[width=0.5\textwidth]{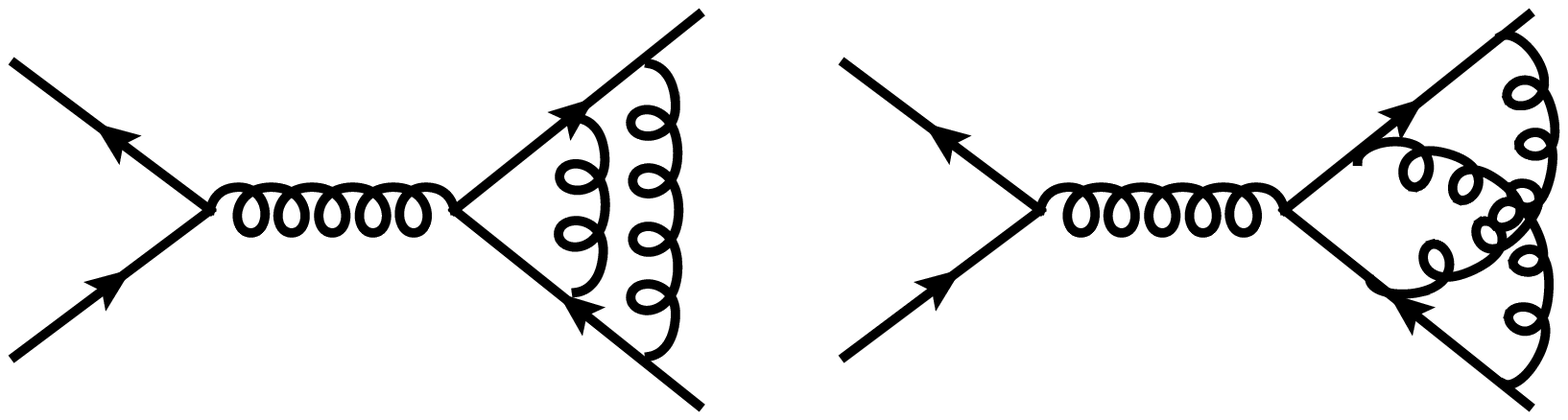} & &
    \includegraphics[width=0.37\textwidth]{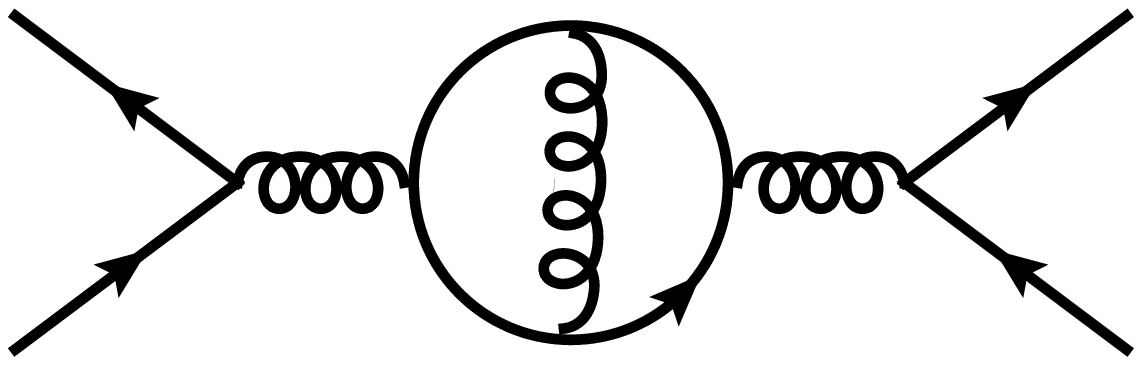}
    \\
    (a) & \hspace*{2em} & (b)
  \end{tabular}
  \caption{\label{fig::vector-current} 
    (a) Examples of two-loop vertex corrections to the 
    colour-octet vector current and (b) the diagram
    responsible for the divergence $n_hT_F(C_A-2C_F)/\epsilon$ in $d_{vv}^{c,(2)}$.}
\end{figure}

All six coefficients are new and not yet present in the literature.
This is also true for the following six matching coefficients obtained from
the annihilation-type diagrams
\begin{align}
c_{s,0}^{c,(2)}=&
\frac{C_F^2}{N_c}\left[-4 \zeta_3-\frac{35}{3}+\frac{\pi ^2}{6}+\frac{40 \log 2}{3}+\frac{7}{9} \pi ^2 \log 2+i\pi \left(\frac{11 \pi ^2}{18}-\frac{20}{3}\right)\right]
+\frac{\pi ^2 C_F n_h T_F}{27 N_c}
\nonumber\\
&
+\frac{C_A C_F}{N_c}\left[\frac{79 \zeta_3}{32}+\frac{751}{108}+\frac{65 \pi ^2}{432}+\frac{11 \log ^22}{9}-\frac{1201 \log 2}{108}-\frac{8}{9} \pi ^2 \log 2
\right.
\nonumber\\
&
\left.
\qquad +i\pi \left(\frac{1201}{216}-\frac{109 \pi ^2}{288}-\frac{11 \log 2}{9}\right)\right]
\nonumber\\
&
+\frac{C_F n_l T_F}{N_c}\left[-\frac{32}{27}+\frac{5 \pi ^2}{27}-\frac{4 \log ^22}{9}+\frac{32 \log 2}{27}+i\pi \left(\frac{4 \log 2}{9}-\frac{16}{27}\right)\right]
+\mathcal{O}(\epsilon)
\,,\nonumber\\
c_{s,1}^{c,(2)}=&
\frac{C_F^2}{N_c}\left[-\frac{3 \zeta_3}{8}-\frac{19}{6}+\frac{4 \pi ^2}{9}+\frac{\log 2}{3}-\frac{1}{18} \pi ^2 \log 2
+i\pi \left(-\frac{1}{6}-\frac{\pi ^2}{72}\right)\right]
+\frac{\pi ^2 C_F n_h T_F}{54 N_c}
\nonumber\\
&
+\frac{C_A C_F}{N_c}\left[\frac{5 \zeta_3}{8}+\frac{535}{216}-\frac{13 \pi ^2}{216}+\frac{11 \log ^22}{18}-\frac{86 \log 2}{27}-\frac{5}{18} \pi ^2 \log 2
\right.
\nonumber\\
&
\left.
+i\pi \left(\frac{43}{27}-\frac{5 \pi ^2}{72}-\frac{11 \log 2}{18}\right)\right]
\nonumber\\
&
+\frac{C_F n_l T_F}{N_c}\left[-\frac{16}{27}+\frac{5 \pi ^2}{54}-\frac{2 \log ^22}{9}+\frac{16 \log 2}{27}+i\pi \left(\frac{2 \log 2}{9}-\frac{8}{27}\right)\right]
+\mathcal{O}(\epsilon)
\,,\nonumber\\
c_{s,2}^{c,(2)}=&
\frac{C_F^2}{N_c}\left[\frac{\zeta_3}{4}+\frac{1}{3}+\frac{\pi ^2}{9}-\frac{2 \log 2}{3}-\frac{1}{9} \pi ^2 \log 2+i\pi \left(\frac{1}{3}-\frac{\pi ^2}{36}\right)\right]
\nonumber\\
&
+\frac{C_A C_F}{N_c}\left[-\frac{3 \zeta_3}{32}-\frac{1}{8}-\frac{\pi ^2}{24}+\frac{\log 2}{4}+\frac{1}{24} \pi ^2 \log 2+i\pi \left(\frac{\pi ^2}{96}-\frac{1}{8}\right)\right]
+\mathcal{O}(\epsilon)
\,,\nonumber\\
c_{o,0}^{c,(2)}=&
C_F^2\left[\frac{\pi ^2}{6 \epsilon}-\frac{51 \zeta_3}{4}+\frac{16 \pi ^2}{3}-\frac{629}{12}+\frac{2}{3} \pi ^2 \log 2+\frac{130 \log 2}{3}+i\pi \left(\frac{23 \pi ^2}{12}-\frac{65}{3}\right)\right]
\nonumber\\
&
+C_A C_F\left[\frac{\pi ^2}{12 \epsilon}+\frac{81 \zeta_3}{4}-\frac{719 \pi ^2}{216}+\frac{1792}{27}+\frac{44 \log ^22}{9}-\frac{43}{18} \pi ^2 \log 2-\frac{1786 \log 2}{27}
\right.
\nonumber\\
&
\left.
+i\pi \left(\frac{893}{27}-\frac{43 \pi ^2}{18}-\frac{44 \log 2}{9}\right)\right]
+C_A n_h T_F\left[\frac{\pi ^2}{16 \epsilon}+\frac{21 \zeta_3}{16}+\frac{5 \pi ^2}{36}+\frac{4613}{648}-\frac{7}{8} \pi ^2 \log 2\right]
\nonumber\\
&
+C_A^2\left[-\frac{\pi ^2}{12 \epsilon}-\frac{33 \zeta_3}{4}-\frac{35 \pi ^2}{72}-\frac{56639}{2592}-\frac{11 \log ^22}{6}+\frac{16}{9} \pi ^2 \log 2+\frac{70 \log 2}{3}
\right.
\nonumber\\
&
\left.
+i\pi \left(\frac{31 \pi ^2}{36}-\frac{35}{3}+\frac{11 \log 2}{6}\right)\right]
+C_F n_h T_F\left[\frac{593 \pi ^2}{864}-\frac{\pi ^2}{8 \epsilon}-\frac{21 \zeta_3}{8}-\frac{277}{36}-\frac{\pi ^2 \log 2}{4}\right]
\nonumber\\
&
+C_F n_l T_F\left[-\zeta_3-\frac{3041}{432}+\frac{20 \pi ^2}{27}-\frac{16 \log ^22}{9}+\frac{373 \log 2}{54}+i\pi \left(\frac{16 \log 2}{9}-\frac{373}{108}\right)\right]
\nonumber\\
&
+C_A n_l T_F\left[\frac{7 \zeta_3}{4}+\frac{3755}{648}-\frac{13 \pi ^2}{36}+\frac{5 \log ^22}{3}-\frac{181 \log 2}{27}+i\pi \left(\frac{181}{54}-\frac{\pi ^2}{12}-\frac{5 \log 2}{3}\right)\right]
\nonumber\\
&
-\frac{64}{81} n_h^2 T_F^2
+n_h n_l T_F^2\left[\frac{32 \log 2}{27}-\frac{80}{81}-\frac{16 i \pi }{27}\right]
\nonumber\\
&
+n_l^2 T_F^2\left[-\frac{25}{81}+\frac{\pi ^2}{9}-\frac{4 \log ^22}{9}+\frac{20 \log 2}{27}+i\pi \left(\frac{4 \log 2}{9}-\frac{10}{27}\right)\right]
+\mathcal{O}(\epsilon)
\,,\nonumber\\
c_{o,1}^{c,(2)}=&
C_A^2\left[\frac{47 \pi ^2}{144}-\frac{43 \zeta_3}{32}-\frac{341}{72}-\frac{11 \log ^22}{12}+\frac{95 \log 2}{18}+\frac{13}{72} \pi ^2 \log 2
+i\pi \left(\frac{31 \pi ^2}{288}-\frac{95}{36}
\right.
\right.
\nonumber\\
&
\left.
\left.
+\frac{11 \log 2}{12}\right)\right]
+C_A C_F\left[\frac{85 \zeta_3}{16}+\frac{1925}{108}-\frac{61 \pi ^2}{54}+\frac{22 \log ^22}{9}-\frac{931 \log 2}{54}-\frac{11}{12} \pi ^2 \log 2
\right.
\nonumber\\
&
\left.
+i\pi \left(\frac{931}{108}-\frac{19 \pi ^2}{48}-\frac{22 \log 2}{9}\right)\right]
+
C_F^2\left[-2 \zeta_3-\frac{40}{3}+\frac{14 \pi ^2}{9}+\frac{8 \log 2}{3}-\frac{4 i \pi }{3}\right]
\nonumber\\
&
+\frac{2}{27} \pi ^2 C_F n_h T_F
+C_F n_l T_F\left[\frac{10 \pi ^2}{27}-\frac{64}{27}-\frac{8 \log ^22}{9}+\frac{64 \log 2}{27}+i\pi \left(\frac{8 \log 2}{9}-\frac{32}{27}\right)\right]
\nonumber\\
&
-\frac{1}{36} \pi ^2 C_A n_h T_F
+C_A n_l T_F\left[\frac{8}{9}-\frac{5 \pi ^2}{36}+\frac{\log ^22}{3}-\frac{8 \log 2}{9}+i\pi \left(\frac{4}{9}-\frac{\log 2}{3}\right)\right]
+\mathcal{O}(\epsilon)
\,,\nonumber\\
c_{o,2}^{c,(2)}=&
C_F^2\left[\frac{3 \zeta_3}{4}+1+\frac{\pi ^2}{3}-2 \log 2-\frac{1}{3} \pi ^2 \log 2+i\pi \left(1-\frac{\pi ^2}{12}\right)\right]
\nonumber\\
&
+C_A C_F\left[-\frac{5 \zeta_3}{8}-\frac{5}{6}-\frac{5 \pi ^2}{18}+\frac{5 \log 2}{3}+\frac{5}{18} \pi ^2 \log 2+i\pi \left(\frac{5 \pi ^2}{72}-\frac{5}{6}\right)\right]
\nonumber\\
&
+C_A^2\left[\frac{5 \zeta_3}{16}+\frac{5}{24}+\frac{17 \pi ^2}{144}-\frac{3 \log 2}{4}-\frac{5}{36} \pi ^2 \log 2+i\pi \left(\frac{3}{8}-\frac{5 \pi ^2}{144}\right)\right]
+\mathcal{O}(\epsilon)
\,.\label{eq:cc-2}
\end{align}
Note that for the annihilation channel, products of two one-loop diagrams
also have to be taken into account. Furthermore, two-loop vertex corrections as
shown in Fig.~\ref{fig::vector-current}(a) contribute to the colour-octet vector
current. After adapting the colour factors, we have cross-checked these
contributions against the explicit results provided Ref.~\cite{Beneke:1997jm}.

In the next subsection we use the results presented above in order to obtain
the four-quark matching coefficients present in ${\cal L}_{\rm NRQCD}$.

\subsection{\label{sub::d_xy}NRQCD four quark coefficients in four dimensions}

In the following we use the expressions from the previous subsection
and apply $[\sigma_i,\sigma_j]=2i\varepsilon_{ijk}\sigma_k$
and $\varepsilon^{jkl}\varepsilon^{jkl'}=2\delta^{ll'}$.
Using Eq.~\eqref{eq::sigma_4} one obtains the following
linear combinations of $c_{s/o,k}$
which provide the matching coefficients present in the NQRCD Lagrange density
of Eq.~(\ref{eq::Ldxy}):
\begin{align}
  d_{ss}&=c_{s,0}+3c_{s,2}\,,\nonumber\\
  d_{vs}&=c_{o,0}+3c_{o,2}\,,\nonumber\\
  d_{sv}&=c_{s,1}-2c_{s,2}\,,\nonumber\\
  d_{vv}&=c_{o,1}-2c_{o,2}\,,
          \label{eq::c2d}
\end{align}
Note that at one-loop
order we have $c_{s/o,2}^{(1)}=0$ and thus the relations are trivial.  The
$\epsilon$-exact one-loop expressions agree with Ref.~\cite{Beneke:2013jia}. Note that in
Ref.~\cite{Pineda:1998kj} a different prescription for $\varepsilon^{ijk}$ in
three
dimensions has been used (cf. discussion between Eqs.~(\ref{eq::sigma_d}) and~(\ref{eq::sigma_4}))
which leads to different relations compared to those in Eq.~(\ref{eq::c2d}).\footnote{At one-loop order
one has $d_{xv}^{\mbox{\tiny\cite{Pineda:1998kj}}} =
(1-\epsilon)d_{xv}^{\mbox{\tiny Eq.(\ref{eq::c2d})}}$ with $x\in\{s,v\}$.}

By denoting the loop corrections as
\begin{align}
  d_{xy} =\sum _{j\geq 0} \pi^2 \left( \frac{\alpha_s^{(n_l)}(m )}{\pi} \right) ^{j+1}
  \left(\frac{\mu^2}{m^2}\right)^{j{\epsilon}}
  d_{xy}^{(j)}
  \,,
  \label{eq:dxy}
\end{align}
the two-loop scattering coefficients are given by
\begin{align}
d_{ss}^{(2)}=&
\frac{C_F^2}{N_c}\left[\frac{3 \zeta_3}{2}-\frac{57}{4}+\frac{55 \pi ^2}{24}\right]
+\frac{C_F n_l T_F}{N_c}\left[\frac{1}{6 \epsilon^2}-\frac{7}{18 \epsilon}-\frac{l_\mu^2}{6} -\frac{\pi ^2}{9}-\frac{19}{9}\right]
+\frac{C_F n_h T_F}{N_c}\left[\frac{\pi ^2}{9}-\frac{20}{27}\right]
\nonumber\\
&
+\frac{C_A C_F}{N_c}\left[-\frac{11}{24 \epsilon^2}+\frac{-\frac{8}{9}-\frac{7 \pi ^2}{24}}{\epsilon}+\frac{11 l_\mu^2}{24}-\frac{119 \zeta_3}{16}+\frac{685 \pi ^2}{288}+\frac{791}{24}+\frac{5}{8} \pi ^2 \log 2\right]
+\mathcal{O}(\epsilon)
\,,\nonumber\\
d_{sv}^{(2)}=&
\frac{C_F^2}{N_c}\left[\frac{\pi ^2}{12 \epsilon}+\frac{15 \zeta_3}{4}+\frac{77 \pi ^2}{48}-\frac{17}{12}-\frac{5}{6} \pi ^2 \log 2\right]
-\frac{5 C_F n_h T_F}{9 N_c}
\nonumber\\
&
+\frac{C_A C_F}{N_c}\left[\frac{7 \pi ^2}{48 \epsilon}+\frac{5 \zeta_3}{2}+\frac{7 \pi ^2}{288}+\frac{1}{36}+\frac{1}{3} \pi ^2 \log 2\right]
+\frac{4 C_F n_l T_F}{9 N_c}
+\mathcal{O}(\epsilon)
\,,\nonumber\\
d_{vs}^{(2)}=&
C_F^2\left[9 \zeta_3-\frac{103}{2}+\frac{227 \pi ^2}{24}\right]
+C_A n_l T_F\left[-\frac{5}{12 \epsilon^2}+\frac{35}{36 \epsilon}+\frac{5 l_\mu^2}{12}+\frac{5 \pi ^2}{18}+\frac{77}{18}\right]
\nonumber\\
&
+C_A C_F\left[-\frac{7}{3 \epsilon^2}+\frac{-\frac{97}{18}-\frac{5 \pi ^2}{6}}{\epsilon}+\frac{11 l_\mu^2}{6}-\frac{65 \zeta_3}{2}+\frac{575 \pi ^2}{144}+\frac{5177}{36}-\pi ^2 \log 2\right]
\nonumber\\
&
+C_A^2\left[\frac{49}{48 \epsilon^2}+\frac{\frac{7}{18}+\frac{\pi ^2}{4}}{\epsilon}-\frac{55}{48} l_\mu^2+\frac{69 \zeta_3}{8}-\frac{275 \pi ^2}{144}-\frac{3161}{72}-\frac{1}{4} \pi ^2 \log 2\right]
\nonumber\\
&
+C_F n_h T_F\left[\frac{4 \pi ^2}{9}-\frac{80}{27}\right]
+C_F n_l T_F\left[\frac{2}{3 \epsilon^2}-\frac{14}{9 \epsilon}-\frac{2}{3} l_\mu^2-\frac{4 \pi ^2}{9}-\frac{76}{9}\right]
\nonumber\\
&
+C_A n_h T_F\left[-\frac{1}{5 \epsilon}-\frac{5 \pi ^2}{18}+\frac{1289}{675}\right]
+\xi\left[\frac{3 C_A^2}{32}+C_A \left(\frac{13}{150}-\frac{1}{20 \epsilon}\right) n_h T_F\right]
+\mathcal{O}(\epsilon)
\,,\nonumber\\
d_{vv}^{(2)}=&
C_F^2\left[\frac{\pi ^2}{4 \epsilon}+\frac{43 \zeta_3}{4}+\frac{155 \pi ^2}{24}-\frac{29}{6}-\frac{23}{6} \pi ^2 \log 2\right]
-\frac{20}{9} C_F n_h T_F
+\frac{16 C_F n_l T_F}{9}
\nonumber\\
&
+C_A C_F\left[\frac{\frac{3 \pi ^2}{8}-\frac{1}{4}}{\epsilon}+\frac{13 \zeta_3}{8}-\frac{137 \pi ^2}{36}+\frac{121}{36}+\frac{59}{12} \pi ^2 \log 2\right]
\nonumber\\
&
+C_A^2\left[\frac{1}{6 \epsilon^2}+\frac{-\frac{11}{72}-\frac{17 \pi ^2}{144}}{\epsilon}-\frac{11}{48} l_\mu^2-\frac{29 \zeta_3}{48}+\frac{275 \pi ^2}{432}-\frac{113}{108}-\frac{29}{24} \pi ^2 \log 2\right]
\nonumber\\
&
+C_A n_h T_F\left[\frac{35}{27}-\frac{\pi ^2}{18}\right]
+C_A n_l T_F\left[-\frac{1}{12 \epsilon^2}+\frac{1}{18 \epsilon}+\frac{l_\mu^2}{12}+\frac{\pi ^2}{18}-\frac{31}{54}\right]
+\mathcal{O}(\epsilon)
\,.\label{eq:d-2}
\end{align}

The relations between $c^c_{s/o,k}$ and $d^c_{xy}$
are also obtained from Eq.~\eqref{eq::sigma_4} and are given by
\begin{align}
  d^c_{ss}&=3c^c_{s,1}-6c^c_{s,2}\,,\nonumber\\
  d^c_{vs}&=3c^c_{o,1}-6c^c_{o,2}\,,\nonumber\\
  d^c_{sv}&=c^c_{s,0}-2c^c_{s,1}+7c^c_{s,2}\,,\nonumber\\
  d^c_{vv}&=c^c_{o,0}-2c^c_{o,1}+7c^c_{o,2}\,.
\end{align}
At tree-level, $c^c_{s/o,1}=c^c_{s/o,2}=0$
and the relations are trivial.

We define the coefficients $d_{xy}^{c,(j)}$ in analogy to
Eq.~\eqref{eq:dxy} and obtain for the one-loop annihilation matching coefficients
\begin{align}
d_{ss}^{c,(1)}=&
\frac{C_F}{N_c} \left[1+\frac{i \pi }{2}-\log 2\right]
+\mathcal{O}(\epsilon)
\,,\nonumber\\
d_{sv}^{c,(1)}=&0
\,,\nonumber\\d_{vs}^{c,(1)}=&
C_A \left[-\frac{3}{2}-\frac{3 i \pi }{4}+\frac{3 \log 2}{2}\right]+C_F (4+2 i \pi -4 \log 2)
+\mathcal{O}(\epsilon)
\,,\nonumber\\
d_{vv}^{c,(1)}=&
-\frac{109 C_A}{36}+4 C_F+\frac{8 n_h T_F}{9}+n_l T_F \left[\frac{5}{9}+\frac{i \pi }{3}-\frac{2 \log 2}{3}\right]
+\mathcal{O}(\epsilon)
\,.\label{eq:dc-1}
\end{align}

The $\epsilon$-exact expressions agree with Ref.~\cite{Beneke:2013jia} and the expanded
expressions with~Ref.~\cite{Pineda:1998kj}.
The two-loop annihilation matching coefficients read
\begin{align}
d_{ss}^{c,(2)}&=
\frac{C_F^2}{N_c}\left[-\frac{21 \zeta_3}{8}-\frac{23}{2}+\frac{2 \pi ^2}{3}+5 \log 2+\frac{1}{2} \pi ^2 \log 2+i\pi \left(\frac{\pi ^2}{8}-\frac{5}{2}\right)\right]
+\frac{\pi ^2 C_F n_h T_F}{18 N_c}
\nonumber\\
&
+\frac{C_A C_F}{N_c}\left[\frac{39 \zeta_3}{16}+\frac{589}{72}+\frac{5 \pi ^2}{72}+\frac{11 \log ^22}{6}-\frac{199 \log 2}{18}-\frac{13}{12} \pi ^2 \log 2+i\pi \left(\frac{199}{36}-\frac{13 \pi ^2}{48}
\right.\right.
\nonumber\\
&
\left.\left.
-\frac{11 \log 2}{6}\right)\right]
+\frac{C_F n_l T_F}{N_c}\left[\frac{5 \pi ^2}{18}-\frac{16}{9}-\frac{2 \log ^22}{3}+\frac{16 \log 2}{9}+i\pi \left(\frac{2 \log 2}{3}-\frac{8}{9}\right)\right]
+\mathcal{O}(\epsilon)
\,,\nonumber\\
d_{sv}^{c,(2)}&=
\frac{C_F^2}{N_c}\left[-\frac{3 \zeta_3}{2}-3+\frac{\pi ^2}{18}+8 \log 2+\frac{1}{9} \pi ^2 \log 2+i\pi \left(\frac{4 \pi ^2}{9}-4\right)\right]
\nonumber\\
&
+\frac{C_A C_F}{N_c}\left[\frac{9 \zeta_3}{16}+\frac{9}{8}-\frac{\pi ^2}{48}-3 \log 2-\frac{1}{24} \pi ^2 \log 2+i\pi \left(\frac{3}{2}-\frac{\pi ^2}{6}\right)\right]
+\mathcal{O}(\epsilon)
\,,\nonumber\\
d_{vs}^{c,(2)}&=
C_F^2\left[-\frac{21 \zeta_3}{2}-46+\frac{8 \pi ^2}{3}+20 \log 2+2 \pi ^2 \log 2+i\pi \left(\frac{\pi ^2}{2}-10\right)\right]
+\frac{2}{9} \pi ^2 C_F n_h T_F
\nonumber\\
&
+C_A C_F\left[\frac{315 \zeta_3}{16}+\frac{2105}{36}-\frac{31 \pi ^2}{18}+\frac{22 \log ^22}{3}-\frac{1111 \log 2}{18}-\frac{53}{12} \pi ^2 \log 2
+i\pi \left(\frac{1111}{36}
\right.\right.
\nonumber\\
&
\left.\left.
-\frac{77 \pi ^2}{48}-\frac{22 \log 2}{3}\right)\right]
+C_A n_l T_F\left[\frac{8}{3}-\frac{5 \pi ^2}{12}+\log ^22-\frac{8 \log 2}{3}+i\pi \left(\frac{4}{3}-\log 2\right)\right]
\nonumber\\
&
+C_A^2\left[-\frac{189 \zeta_3}{32}-\frac{371}{24}+\frac{13 \pi ^2}{48}-\frac{11 \log ^22}{4}+\frac{61 \log 2}{3}+\frac{11}{8} \pi ^2 \log 2
\right.
\nonumber\\
&
\left.
+i\pi \left(-\frac{61}{6}+\frac{17 \pi ^2}{32}+\frac{11 \log 2}{4}\right)\right]
-\frac{1}{12} \pi ^2 C_A n_h T_F
\nonumber\\
&
+C_F n_l T_F\left[-\frac{64}{9}+\frac{10 \pi ^2}{9}-\frac{8 \log ^22}{3}+\frac{64 \log 2}{9}+i\pi \left(\frac{8 \log 2}{3}-\frac{32}{9}\right)\right]
+\mathcal{O}(\epsilon)
\,,\nonumber\\
d_{vv}^{c,(2)}&=
C_F^2\left[\frac{\pi ^2}{6 \epsilon}-\frac{7 \zeta_3}{2}+\frac{41 \pi ^2}{9}-\frac{75}{4}-\frac{5}{3} \pi ^2 \log 2+24 \log 2+i\pi \left(\frac{4 \pi ^2}{3}-12\right)\right]
\nonumber\\
&
+C_A C_F\left[\frac{\pi ^2}{12 \epsilon}+\frac{21 \zeta_3}{4}-\frac{217 \pi ^2}{72}+\frac{224}{9}+\frac{25}{18} \pi ^2 \log 2-20 \log 2+i\pi \left(10-\frac{10 \pi ^2}{9}\right)\right]
\nonumber\\
&
+C_A^2\left[-\frac{\pi ^2}{12 \epsilon}-\frac{27 \zeta_3}{8}-\frac{5 \pi ^2}{16}-\frac{28307}{2592}+\frac{4}{9} \pi ^2 \log 2+\frac{271 \log 2}{36}+i\pi \left(\frac{29 \pi ^2}{72}-\frac{271}{72}\right)\right]
\nonumber\\
&
+C_F n_h T_F\left[-\frac{\pi ^2}{8 \epsilon}-\frac{21 \zeta_3}{8}+\frac{155 \pi ^2}{288}-\frac{277}{36}-\frac{1}{4} \pi ^2 \log 2\right]
-\frac{64}{81} n_h^2 T_F^2
\nonumber\\
&
+C_F n_l T_F\left[-\zeta_3-\frac{331}{144}+\frac{13 \log 2}{6}-\frac{13 i \pi }{12}\right]
+n_h n_l T_F^2\left[\frac{32 \log 2}{27}-\frac{80}{81}-\frac{16 i \pi }{27}\right]
\nonumber\\
&
+C_A n_h T_F\left[\frac{\pi ^2}{16 \epsilon}+\frac{21 \zeta_3}{16}+\frac{7 \pi ^2}{36}+\frac{4613}{648}-\frac{7}{8} \pi ^2 \log 2\right]
\nonumber\\
&
+C_A n_l T_F\left[\frac{7 \zeta_3}{4}+\frac{2603}{648}-\frac{\pi ^2}{12}+\log ^22-\frac{133 \log 2}{27}+i\pi \left(\frac{133}{54}-\frac{\pi ^2}{12}-\log 2\right)\right]
\nonumber\\
&
+n_l^2 T_F^2\left[-\frac{25}{81}+\frac{\pi ^2}{9}-\frac{4 \log ^22}{9}+\frac{20 \log 2}{27}+i\pi \left(\frac{4 \log 2}{9}-\frac{10}{27}\right)\right]
+\mathcal{O}(\epsilon)
\,.\label{eq:dc-2}
\end{align}

Note that all two-loop coefficients are $\xi$ independent except
$d_{vs}^{(2)}$. In fact, the gauge parameter dependence cancels in the
combination $(\alpha_s/\pi) c_D^{(2)} + d_{vs}^{(2)}$ which enters
physical quantities.

The imaginary parts of $d_{ss}^{c,(2)}$, 
$d_{vs}^{c,(2)}$, and $d_{vv}^{c,(2)}$ are calculated 
in the context of the heavy quarkonium inclusive decays~\cite{Vairo:2003gh},
and our results agree with the literature.

All the matching coefficients from the annihilation process are finite after
the UV renormalization except $d_{vv}^{c,(2)}$.  The remaining divergences
originate from diagrams shown in Fig.~\ref{fig::vector-current}(a) and~(b).
They are well studied in the literature~\cite{Beneke:1997zp} where it is shown
that the divergences from the purely hard regions, which are contained in our
expressions, are canceled against contributions from the potential region.  We
have confirmed this cancellation for the contribution from
Fig.~\ref{fig::vector-current}(b) where explicit results for the different regions
are given in Ref.~\cite{Beneke:1997zp}.


\section{\label{sec::concl}Conclusions and outlook}

In this paper we compute two-loop corrections to the matching coefficients
$d_{ss}$, $d_{sv}$, $d_{vs}$, $d_{vv}$,
$d^c_{ss}$, $d^c_{sv}$, $d^c_{vs}$ and $d^c_{vv}$ of the operators in the NRQCD
Lagrange density involving four heavy quarks. We carefully discuss the
treatment of the Pauli matrices in a non-integer number of dimensions which leads to an
enlargement of the basis and six (instead of four) two-loop coefficients in
intermediate steps (see Section~\ref{sub::c_xy}). The results for $d_{xy}$
and $d^c_{xy}$,
which are obtained after using the usual commutation relations between the
Pauli matrices, are given in Section~\ref{sub::d_xy}.

Our calculation is performed in the covariant $R_\xi$ gauge with a general gauge
parameter $\xi$. One observes that starting from two loops the coefficient
$d_{vs}$ is $\xi$ dependent which arises from our non-minimal choice of the
operator basis in $\mathcal{L}_{\text{NRQCD}}$.  We check the $\xi$ dependence
by computing two-loop corrections to the heavy-quark-gluon vertex
functions. We extract the related matching coefficients, in particular $c_D$,
and show that the combination $(\alpha_s/\pi) c_D^{(2)} + d_{vs}^{(2)}$ is
independent of $\xi$.  Note that in Feynman gauge the one-loop results
$c_D^{(1)}$ and $d_{vs}^{(1)}$ are individually $\xi$ independent. However,
the gauge dependence can be observed by comparing to the results in Coulomb
gauge~\cite{Pineda:2001ra}.

The results obtained in this paper enter as building blocks various
physical quantities involving two slowly moving heavy quarks at the N$^3$LL
and N$^4$LO accuracy.

The annihilation channel only contributes to the case where the two heavy
quarks in ${\cal L}_{\phi\chi}$ (cf. Eq.~(\ref{eq::Ldxy})) have
the same flavour. On the other hand, for different quark flavours the matching
coefficients $d_{xy}$ receive contributions only from the scattering channel.
We use the same mass for quarks and anti-quarks and provide only results for this equal-mass case.
A possible next step would thus
be the extension of our calculation of the scattering contribution
to the case of different quark masses. A further next step is the computation
of two-loop corrections to the matching coefficient of the operator
with two heavy and two light quarks usually denoted by $c_1^{hl}$ (see, e.g.,
Ref.~\cite{Anzai:2018eua}).



\section*{Acknowledgements}
We would like to thank Alexander Penin, Jan Piclum and Antonio Pineda for many
useful discussions and communications. We thank Florian Herren for technical
help in connection to the partial fraction decomposition and Joshua
Davies for carefully reading the manuscript.  This research was
supported by the Deutsche Forschungsgemeinschaft (DFG, German Research
Foundation) under grant 396021762 --- TRR 257 ``Particle Physics Phenomenology
after the Higgs Discovery'' and the Graduiertenkolleg
``Elementarteilchenphysik bei h\"ochster Energie und h\"ochster Pr\"azision''.


\begin{appendix}


\section{\label{app::master}Master integrals}

In this appendix we collect analytic results for the master integrals which we
need for the computation of the matching coefficient. Most of them are already
needed for two-loop matching coefficients between QCD and NRQCD of the vector,
axial-vector, scalar and pseudo-scalar
currents~\cite{Czarnecki:1997vz,Beneke:1997jm,Kniehl:2006qw} and the integrals have
been studied in the literature~\cite{Piclum:2007an} (see
also Refs.~\cite{Fleischer:1999tu,Anastasiou:2006hc}). Note, however, that for $I_2^e$ 
the $\epsilon$ expansion was not sufficiently deep and the $\epsilon^0$ was
only known numerically. Furthermore $I_2^g$ was (to our
knowledge) not available in the literature.

The master integrals are defined as (cf. Fig.~\ref{fig::masters})
\begin{align}
I_1^a=
&\frac{\mathcal{N}}{m^2}
\int \frac{\mathrm{d}^d k}{i\pi^{d/2}}
\frac{-1}{k^2-m^2}
\,,\nonumber\\
I_1^b=
&\mathcal{N}\int \frac{\mathrm{d}^d k}{i\pi^{d/2}}
\frac{-1}{k^2}
\frac{-1}{(k+2p)^2}
\,,\nonumber\\
I_2^a=
&\frac{\mathcal{N}^2}{m^2}
\int \frac{\mathrm{d}^d k}{i\pi^{d/2}}
\frac{\mathrm{d}^d \ell}{i\pi^{d/2}}
\frac{-1}{k^2-m^2}
\frac{-1}{\ell^2}
\frac{-1}{(k+\ell+p)^2}
\,,\nonumber\\
I_2^b=
&\frac{\mathcal{N}^2}{m^2}
\int \frac{\mathrm{d}^d k}{i\pi^{d/2}}
\frac{\mathrm{d}^d \ell}{i\pi^{d/2}}
\frac{-1}{k^2-m^2}
\frac{-1}{\ell^2-m^2}
\frac{-1}{(k+\ell+p)^2-m^2}
\,,\nonumber\\
I_2^c=
&\frac{\mathcal{N}^2}{m^2}
\int \frac{\mathrm{d}^d k}{i\pi^{d/2}}
\frac{\mathrm{d}^d \ell}{i\pi^{d/2}}
\frac{-1}{k^2}
\frac{-1}{\ell^2-m^2}
\frac{-1}{(k+\ell+2p)^2-m^2}
\,,\nonumber\\
I_2^d=
&\frac{\mathcal{N}^2}{m^2}
\int \frac{\mathrm{d}^d k}{i\pi^{d/2}}
\frac{\mathrm{d}^d \ell}{i\pi^{d/2}}
\frac{-1}{k^2}
\frac{-1}{\ell^2}
\frac{-1}{(k+\ell+2p)^2}
\,,\nonumber\\
I_2^e=
&\mathcal{N}^2\int \frac{\mathrm{d}^d k}{i\pi^{d/2}}
\frac{\mathrm{d}^d \ell}{i\pi^{d/2}}
\frac{-1}{k^2-m^2}
\frac{-1}{(\ell+p)^2}
\frac{-1}{(\ell-p)^2}
\frac{-1}{(k+\ell)^2}
\,,\nonumber\\
I_2^f=
&\mathcal{N}^2m^2\int \frac{\mathrm{d}^d k}{i\pi^{d/2}}
\frac{\mathrm{d}^d \ell}{i\pi^{d/2}}
\frac{-1}{(\ell+p)^2}
\frac{-1}{(\ell-p)^2}
\frac{-1}{(k+p)^2-m^2}
\frac{-1}{(k-p)^2-m^2}
\frac{-1}{(k+\ell)^2-m^2}
\,,\nonumber\\
I_2^g=
&\mathcal{N}^2m^2\int \frac{\mathrm{d}^d k}{i\pi^{d/2}}
\frac{\mathrm{d}^d \ell}{i\pi^{d/2}}
\frac{-1}{k^2-m^2}
\frac{-1}{(k+p)^2}
\frac{-1}{\ell^2-m^2}
\frac{-1}{(\ell+p)^2}
\frac{-1}{(k+\ell)^2}
\,,
\end{align}
where $\mathcal{N}=(\mu^2 e^{\gamma_\mathrm{E}})^\epsilon$.
We normalize the master integrals such that 
they have the mass dimension zero.
Our results read
\begin{align}
I_1^a=&\left(\frac{\mu^2}{m^2}e^{\gamma_E}\right)^{\epsilon}
\Gamma (\epsilon -1)
\,,\nonumber\\
I_1^b=&\left(\frac{\mu^2}{m^2}e^{\gamma_E}\right)^{\epsilon}
\frac{e^{i\pi\epsilon}}{4^\epsilon}
\frac{\Gamma (1-\epsilon)^2\Gamma(\epsilon)}
{\Gamma(2-2\epsilon)}
\,,\nonumber\\
I_2^a=&\left(\frac{\mu^2}{m^2}e^{\gamma_E}\right)^{2\epsilon}
\frac{\Gamma(1-\epsilon)^2 \Gamma(\epsilon)}{\Gamma(2-2 \epsilon)} 
\frac{\Gamma(2 \epsilon-1) \Gamma(3-4 \epsilon)}{\Gamma(3-3 \epsilon)}
\,,\nonumber\\
I_2^b=&\left(\frac{\mu^2}{m^2}\right)^{2\epsilon}
\left\{
-\frac{3}{2 \epsilon^{2}}
-\frac{17}{4 \epsilon}
-\frac{59}{8}
-\frac{\pi^{2}}{4}
-\left(\frac{65}{16}
+\frac{49}{24} \pi^{2}-\zeta_{3}\right) \epsilon
\right.
\nonumber\\
&
\left.
-\left(-\frac{1117}{32}
+\frac{475}{48} \pi^{2}
-8 \pi^{2} \log  2
+\frac{151}{6} \zeta_{3}
+\frac{7}{240} \pi^{4}\right) \epsilon^{2}
+\mathcal{O}(\epsilon^3)
\right\}
\,,\nonumber\\
I_2^c=&\left(\frac{\mu^2}{m^2}\right)^{2\epsilon}
\left\{
-\frac{1}{\epsilon^{2}}
-\frac{2}{\epsilon}
+\frac{1}{2}
-\frac{11}{12} \pi^{2}
\right.
\nonumber\\
&
\left.
-\left(-\frac{85}{4}
+\frac{17}{24} \pi^{2}
+\frac{3}{2} \pi^{2} \log  2
+\frac{181}{12} \zeta_{3}\right) \epsilon
+\mathcal{O}(\epsilon^2)
\right\}
\,,\nonumber\\
I_2^d=&-4
\left(\frac{\mu^2}{m^2}e^{\gamma_E}\right)^{2\epsilon}
\frac{e^{2i\pi\epsilon}}{4^{2\epsilon}}
\frac{\Gamma (1-\epsilon)^3\Gamma(2\epsilon-1)}
{\Gamma(3-3\epsilon)}
\,,\nonumber\\
I_2^e=&\left(\frac{\mu^2}{m^2}\right)^{2\epsilon}
\left\{
\frac{1}{2\epsilon^{2}}
+\frac{1}{\epsilon}
\left(\frac{5}{2}-2\log  2\right)
+\frac{19}{2}-\frac{13 \pi ^2}{12}+4 \log ^2 2 -8 \log  2
\right.
\nonumber\\
&+\epsilon 
\left(\frac{65}{2}
-\frac{77 \zeta (3)}{6}
-\frac{47 \pi ^2}{12}
-\frac{16 \log  ^3 2}{3}
+16 \log  ^2 2-24 \log  2
+\frac{13}{3} \pi ^2 \log  2\right)
\nonumber\\
&
\left.
+i\left[
\frac{\pi}{\epsilon}
+4\pi (1-\log  2)
+\epsilon \left(12 \pi -\frac{\pi ^3}{3}+8 \pi  \log  ^2 2-16 \pi  \log  2\right)
\right]
+\mathcal{O}(\epsilon^2)
\right\}
\,,\nonumber\\
I_2^f=&
\left(\frac{\mu^2}{m^2}\right)^{2\epsilon}
\left\{
\frac{1}{2} \pi^{2} \log  2-\frac{21}{8} \zeta_{3}
+i \frac{1}{8} \pi^{3}+\mathcal{O}(\epsilon)
\right\}
\,,\nonumber\\
I_2^g=&
\left(\frac{\mu^2}{m^2}\right)^{2\epsilon}
\left\{
\frac{2}{3} \pi^{2} \log  2-\frac{3}{2} \zeta_{3}
+i \frac{1}{6} \pi^{3}+\mathcal{O}(\epsilon)
\right\}
\label{eq:MasterIntegrals}
\,.
\end{align}
For the integral $I_2^e$ we derive 
a Mellin-Barnes representation with non-zero parameter $\epsilon$
and use \texttt{MB.m}~\cite{Czakon:2005rk} 
to analytically continue to $\epsilon\to 0$.
The resulting (at most) two-dimensional Mellin-Barnes integrals
are reduced to one-dimensional Mellin-Barnes integrals
with a help of the generalized Barnes lemma~\cite{Davies:2018ood,Mishima:2018olh}.
The one-dimensional integrals can be evaluated numerically with a very high precision,
and we apply the PSLQ algorithm~\cite{PSLQ}
to obtain the analytic results.

Using the Mellin-Barnes method for $I_2^g$ leads to a complicated
four-dimensional Mellin-Barnes integral, and we adopt a different strategy for
its computation. Note that $I_2^g$ is a finite integral and we require only
the $\epsilon^0$ term.  This means we can set $\epsilon=0$ from the very
beginning of our computation. We use the Lee-Pomeransky
representation~\cite{Lee:2013hzt} which turns out to be useful since the
integrand is now a simple rational function.  We can perform most of the
integrations analytically and remain only with a two-dimensional integral with
good covergence properties. Thus, numerical integration leads to sufficiently
high precision such that the PSLQ algorithm can be applied.  We cross-check
all master integrals with the help of \texttt{FIESTA}~\cite{Smirnov:2015mct}.


\section{\label{app::d_i}Gluon field redefinition}

In Ref.~\cite{Pineda:2011dg} the NRQCD Lagrange density has been defined such
that the kinetic term of the gluon field has a canonical normalization which
has been achieved by a redefinition of the gluon field.  The procedure is
presented in Ref.~\cite{Pineda:2000sz}.  As a consequence the constants $d_1$
and $d_2$ appear on the right-hand side of the formula for $c_D$ in
Eq.~(\ref{eq::cD}). In this section we provide analytic expressions for $d_1$
and $d_2$ up to two-loop order.

Our starting point is the following Lagrange density which describes 
the interaction of the heavy quarks with a gluon before the redefinition of
the gluon field
\begin{eqnarray}
  \delta{\cal L}_{\rm NRQCD}^g
  \!\!&\!=\!&\!\!
      -\frac{d_1}{4}G^{a}_{\mu \nu}G^{a\mu\nu}
      +\frac{d_2}{m^2}G^{a}_{\mu \nu}D^2G^{a\mu\nu}
      +\frac{d_3}{m^2}g_sf^{abc}G^{a}_{\mu \nu}G^{b\mu}_{\phantom{b\mu}
      \alpha}G^{c\nu \alpha}+O\bigg(\frac{1}{m^4}\bigg)\,,
      \label{eq::Lg}
\end{eqnarray}
where $G_{\mu\nu}$ is the gluon field strength tensor and $a,b,c$ are colour
indices.  The matching coefficients $d_1$ and $d_2$ can be computed from the
hard contribution of the gluon two-point function.  For convenience we provide
the results  which we parametrize by
\begin{align}
  d_{i}=\sum _{j\geq 0} \left( \frac{\alpha_s^{(n_l+n_h)} (m)}{\pi} \right) ^j 
  \left(\frac{\mu^2}{m^2}\right)^{j}
  d_{i}^{(j)}
  \,,
\end{align}
and $d_{1}^{(0)}=1$, $d_{2}^{(0)}=0$.
Up to two-loop order our results read
\begin{align}
d_1^{(1)}=&
\frac{1}{3} n_h T_F l_\mu
+\mathcal{O}(\epsilon)
\,,\nonumber\\
d_2^{(1)}=&
\frac{n_h T_F}{60}
+\mathcal{O}(\epsilon)
\,,\nonumber\\
d_1^{(2)}=&
C_F n_h T_F\left[\frac{l_\mu}{4}+\frac{15}{16}\right]
+C_A n_h T_F\left[-\frac{11}{36} l_\mu^2+\frac{5 l_\mu}{12}-\frac{2}{9}\right]
+\frac{1}{9} n_h^2 T_F^2 l_\mu^2
+\frac{1}{9} n_h n_l T_F^2 l_\mu^2
+\mathcal{O}(\epsilon)
\,,\nonumber\\
d_2^{(2)}=&
\frac{41 C_F n_h T_F}{648}
+C_A n_h T_F\left[\frac{1}{960 \epsilon}-\frac{4957}{259200}\right]
+\frac{1}{180} n_h^2 T_F^2 l_\mu
+\frac{n_h n_l T_F^2}{180 \epsilon}
\nonumber\\
&
+\xi\left[C_A \left(\frac{1}{320 \epsilon}-\frac{13}{2400}\right) n_h T_F\right]
+\mathcal{O}(\epsilon)
\,.\label{eq:d1d2-2}
\end{align}
Note that the external gluon fields have been renormalized in the
$\overline{\rm MS}$ scheme.

It is common practice to perform a redefinition of the gluon field as
\begin{align}
  A_{\mu} \to A_{\mu}+\frac{2d_2}{d_1 m^2} \left[D^{\alpha}, G_{\alpha \mu}\right]\,,
\end{align}
which eliminates the second term in Eq.~\eqref{eq::Lg}.
A subsequent rescaling of the form
\begin{align}
  A_{\mu} \to \frac{1}{\sqrt{d_1}} A_{\mu}\,,
  \label{eq::A_rescale}
\end{align}
leads to the canonical factor ``$-1/4$'' in the first term
of Eq.~\eqref{eq::Lg}.


\section{\label{app::cFcS}Results for $c_F$ and $c_S$}

In this appendix we provide analytic results for $c_F$ and $c_S$ up to two
loops. Our results read
\begin{align}
c_F^{(1)}=&
C_A \left[\frac{1}{4 \epsilon}+\frac{1}{2}\right]+\frac{C_F}{2}
+\mathcal{O}(\epsilon)
\,,\nonumber\\
c_S^{(1)}=&
C_A \left[\frac{1}{2 \epsilon}+1\right]+C_F
+\mathcal{O}(\epsilon)
\,,\nonumber\\
c_F^{(2)}=&
C_A^2\left[-\frac{1}{12 \epsilon^2}+\frac{35}{144 \epsilon}+\frac{11 l_\mu^2}{96}-\frac{\zeta_3}{8}-\frac{65 \pi ^2}{576}+\frac{859}{432}+\frac{1}{12} \pi ^2 \log 2\right]
+C_F n_h T_F\left[\frac{119}{36}-\frac{\pi ^2}{3}\right]
\nonumber\\
&
-\frac{25}{36} C_F n_l T_F
+C_A n_h T_F\left[\frac{5 \pi ^2}{72}-\frac{149}{216}\right]
+C_A n_l T_F\left[\frac{1}{24 \epsilon^2}-\frac{13}{144 \epsilon}-\frac{1}{24} l_\mu^2-\frac{\pi ^2}{36}-\frac{299}{432}\right]
\nonumber\\
&
+ C_F^2\left[\frac{3 \zeta_3}{4}-\frac{31}{16}+\frac{5 \pi ^2}{12}-\frac{1}{2} \pi ^2 \log 2\right]
+C_A C_F\left[\frac{1}{8 \epsilon}-\frac{\zeta_3}{8}+\frac{\pi ^2}{12}+\frac{341}{144}+\frac{\pi ^2 \log 2}{12} \right]
+\mathcal{O}(\epsilon)
\,,\nonumber\\
c_S^{(2)}=&
C_A^2\left[-\frac{1}{6 \epsilon^2}+\frac{35}{72 \epsilon}+\frac{11 l_\mu^2}{48}-\frac{\zeta_3}{4}-\frac{65 \pi ^2}{288}+\frac{859}{216}+\frac{1}{6} \pi ^2 \log 2\right]
+C_F n_h T_F\left[\frac{119}{18}-\frac{2 \pi ^2}{3}\right]
\nonumber\\
&
-\frac{25}{18} C_F n_l T_F
+C_A n_h T_F\left[\frac{5 \pi ^2}{36}-\frac{149}{108}\right]
+C_A n_l T_F\left[\frac{1}{12 \epsilon^2}-\frac{13}{72 \epsilon}-\frac{1}{12} l_\mu^2-\frac{\pi ^2}{18}-\frac{299}{216}\right]
\nonumber\\
&
+ C_F^2\left[\frac{3 \zeta_3}{2}-\frac{31}{8}+\frac{5 \pi ^2}{6}-\pi ^2 \log 2\right]
+C_A C_F\left[\frac{1}{4 \epsilon}-\frac{\zeta_3}{4}+\frac{\pi ^2}{6}+\frac{341}{72}+\frac{1}{6} \pi ^2 \log 2\right]
+\mathcal{O}(\epsilon)
\,.\label{eq:cFcS-2}
\end{align}

The one-loop results agree with 
Refs.~\cite{Manohar:1997qy} and~\cite{Beneke:2013jia};
the two-loop results are new.


\end{appendix}


\end{document}